\documentclass[10pt]{article}
\usepackage[T1]{fontenc}
\usepackage[a4paper]{geometry}
\usepackage{amsmath}
\usepackage{amssymb}
\usepackage{gr
aphicx}
\usepackage{setspace}
\usepackage{esint}
\usepackage{xcolor}
\makeatletter

\providecommand{\tabularnewline}{\\}

\def\bg #1{\mbox{\boldmath{$#1$}}}

\@ifundefined{showcaptionsetup}{}{%
 \PassOptionsToPackage{caption=false}{subfig}}
\usepackage{subfig}
\makeatother

\begin{document}

\title{\textbf{\large{}COMPUTATIONAL HOMOGENIZATION OF FIBROUS PIEZOELECTRIC
MATERIALS}}

\maketitle
\begin{center}
Claudio Maruccio$^{1}$, Laura De Lorenzis$^{2}$, Luana Persano$^{3}$,
Dario Pisignano$^{3,4}$
\par\end{center}

\begin{center}
{\small{}$^{1}$ Dipartimento di Ingegneria dell'Innovazione, Università
del Salento, Via Monteroni, Lecce, Italy. E-mail: claudio.maruccio@unisalento.it}
\par\end{center}{\small \par}

\begin{center}
{\small{}$^{2}$ Institut für Angewandte Mechanik, Technische Universität
Braunschweig, Bienroder Weg, Braunschweig, Germany. E-mail: l.delorenzis@tu-braunschweig.de}
\par\end{center}{\small \par}

\begin{center}
{\small{}$^{3}$ National Nanotechnology Laboratory, CNR-Istituto
Nanoscienze, via Arnesano, Lecce, Italy. E-mail: luana.persano@nano.cnr.it}
\par\end{center}{\small \par}

\begin{center}
{\small{}$^{4}$ Dipartimento di Matematica e Fisica ``E. De Giorgi'',
Università del Salento, via Monteroni, Lecce, Italy. E-mail: dario.pisignano@unisalento.it}
\par\end{center}{\small \par}


\paragraph{Summary}

\textit{Flexible piezoelectric devices made of polymeric materials
are widely used for micro- and nano-electro-mechanical systems. In
particular, numerous recent applications concern energy harvesting.
Due to the importance of computational modeling to understand the
influence that microscale geometry and constitutive variables exert
on the macroscopic behavior, a numerical approach is developed here
for multiscale and multiphysics modeling of thin piezoelectric sheets
made of aligned arrays of polymeric nanofibers, manufactured by electrospinning. At the microscale, the representative volume element consists in piezoelectric polymeric
nanofibers, assumed to feature a piezoelastic behavior and subjected to electromechanical contact constraints.  The latter are incorporated into the virtual work equations by formulating suitable electric, mechanical and coupling potentials and the constraints are enforced by using the penalty method. From the solution of the micro-scale boundary value problem, a suitable scale transition procedure leads to identifying the performance of
a macroscopic thin piezoelectric shell element. }

\section{{\normalsize{}Introduction}}

The discovery of appreciable piezoelectricity on ceramic and polymeric
materials such as ZnO and Polyvinylidene fluoride (PVDF) has led to
the development of a series of nanowire-based piezoelectric nanogenerators
\cite{key-1,key-100001,100002}. These piezoelectric devices are attractive
for several technological applications, most notably for mechanical
energy harvesting and pressure/force sensors. The impact of these
technologies e.g. for powering small electronic devices and, in perspective,
coupling with sensing and photonic platforms, is significant and foreseeably
going to grow.
\color{black}
In particular, PVDF is a polymeric piezoelectric material with good piezoelectric and mechanical properties. For comparison, the most important piezoelectric coefficients are for ZnO: $d_{33}$=12 pC/N, $d_{31}$=-4.7 pC/N and $d_{15}$=-12 pC/N, and for bulk PVDF  $d_{33}$=-30 pC/N, $d_{31}$=23 pC/N, $d_{15}$=0 pC/N.
Recent experimental results showed that the piezoelectric coefficient $d_{33}$ of PVDF in the form of nanofibers is around -40 pC/N, i.e. is higher $($in absolute value$)$ than that of ZnO and bulk PVDF. In particular, considering PVDF in the form of nanofiber arrays an increase up to 20-30$\%$ is found as a function of the alignment of the fibers at the microscale \cite{key-100003,key-100004,key-903}.
\color{black}
Moreover, the high flexibility
typical of polymers allows very high strains to be applied to a PVDF
sheet or filament, thus high piezoelectric potentials can be expected.
Electrospinning technologies, which exploit the elongation of electrified
jets to produce polymer nanofibers, are attracting an increasing attention
in this framework, because they are especially effective, cheap and
allow good amounts of functional fibers to be realized in continuous
runs \cite{key-2001,key-2002,key-2003,key-2004}. In particular, a
few electrospinning methods were applied to piezoelectric polymers
to realize active, eventually self-poled piezoresponsive fibers. For instance, far-field electrospinning was recently used \cite{key-2,key-903}
to produce piezoelectric thin sheets of PVDF nanofibers aligned in
uniaxial arrays. This approach allows macroscopic samples (sheets)
of aligned nanofibers to be obtained, as well as different microscale
fiber architectures to be produced which may vary in a controlled
way from a completely randomly oriented to a fully aligned ensemble
depending on the process parameters and specific experimental set-up, as shown in Fig. \ref{fig:1}.
\color{black}
During the fabrication process we observed that welding takes place between adjacent fibers which leads the assembly of nanofibers to behave as a structure.
\color{black}
The stretching force and a strong electric
field applied between the nozzle and the collecting substrate surface
pole the PVDF nanofibers into the $\beta$ phase. The resulting PVDF
sheet can then be directly applied between two electrodes on a flexible
plastic substrate for piezoelectric output measurement \cite{key-2,key-903}.
As shown in Fig. \ref{fig:1}e, the material which macroscopically
appears in the form of thin sheets is actually made of aligned arrays
of polymeric nanofibers. This becomes evident when looking at it one
scale below (termed ``microscale'' in the following).

Due to the importance of understanding the influence that microscale
geometry and constitutive variables exert on the macroscopic behavior,
the objective of this work is to develop a multiphysics multiscale
procedure to obtain the macroscopic piezoelastic behavior of the thin
sheets based on the microscale features of the constituent material.
In order to predict the macroscale properties of materials and devices
featuring heterogeneous properties at the lower scale(s), several
analytical and computational multiscale approaches have been developed
in the past years \cite{key-5,key-6} to overcome the prohibitive
computational expense required for an explicit description of the
lower-scale features. Although most of these efforts have been devoted
to continuum mechanics \cite{key-7,key-8}, some applications to multiphysics
problems are also available. A few of these focus specifically on
electromechanically coupled problems such as in the case of piezoelectricity
\cite{key-27}, and are based either on analytical approaches \cite{key-10,key-11}
or on finite element analyses \cite{key-12,key-13}. Although several
macroscale formulations were developed for piezoelectric shell elements
\cite{key-14,key-15,key-16}, computational homogenization of shells
is only recently receiving major attention \cite{key-17}. Existing
approaches model the representative volume element (RVE) as an ordinary
3D continuum, described by the standard kinematic, equilibrium and
constitutive equations. In-plane homogenization is combined with through-thickness
integration to obtain the macroscopic generalized stress and moment
resultants \cite{key-18}. To the best of our knowledge, such approaches
have not yet been proposed for piezoelectric shells.

In this paper, we present a computational homogenization procedure
which derives macroscopic non-linear constitutive laws for piezoelectric
shells made of aligned arrays of polymeric nanofibers from the detailed
description of the microscale. The motivation is to describe the behavior
of the PVDF sheets mentioned earlier and, in a subsequent phase, to
optimize their macroscopic piezoelectric response by tailoring their
microscale features. The concept is schematically illustrated in Fig.
\ref{fig:6}.

This paper is organized as follows: Section 2 describes the kinematic
behavior of a piezoelectric shell, where displacements and electric
potential are the independent fields. In Section 3 a microscale RVE
element is defined and the \color{black} kinematically nonlinear theory of piezoelasticity
\color{black} is briefly reviewed along with its finite element formulation at the microscale. \color{black} The implementation of suitable frictionless electromechanical contact
elements using smoothing techniques is described and details regarding
the linearization of the finite element equations at the micro scale
are also provided. Section 4 describes the transition between the
micro- and macro scales. Finally, in Section 5 the RVE geometry is
analyzed to determine the effective material properties of the macroscopic
shell. Advanced symbolic computational tools available in the AceGen/AceFEM
finite element environment within Mathematica \cite{key-19,key-20}
are used throughout this work, with the advantage that the tasks related
to the finite element implementation, including linearization of the
non-linear governing equations, are largely automated.

\section{{\normalsize{}Macroscale: shell kinematics}}
In this section we describe the kinematics assumed for the macroscale
piezoelectric shell, Fig. \ref{fig:7}, \color{black} following \cite{key-14}, which in turn is largely based on Naghdi's theory for the mechanical part \cite{key-901,key-904}. For more details, see the original paper [27]. The Green-Lagrange strains
and the electric field are derived in convective coordinates. The
parameter $\xi^{3}\in\left[-\frac{1}{2},\frac{1}{2}\right]$ is defined
as the thickness parametric coordinate and $\xi^{\alpha}$ with $\alpha=1,2$
are the in-plane parametric coordinates of the shell middle surface.
Thus $\left(\xi^{1},\xi^{2}\right)\in\mathcal{A}\subset\mathbb{R}^{2}$,
with $\mathcal{A}$ as the domain of the shell middle surface parameterization.\color{black}
We further introduce the convention that indices in Greek letters take
the values $1,2$ whereas indices in Latin letters take the values
$1,2,3$. Moreover, partial derivatives are denoted as follows
\begin{equation}
(\circ)_{,\alpha}=\frac{\partial(\circ)}{\partial\xi^{\alpha}},\:\alpha=1,2;\quad(\circ)_{,i}=\frac{\partial(\circ)}{\partial\xi^{i}},\: i=1,2,3\label{eq12}
\end{equation}

\subsection{{\normalsize{}Mechanical field}}

The position vector in the reference (initial) shell configuration
is defined as :
\begin{equation}
\mathbf{X}\left(\xi^{1},\xi^{2},\xi^{3}\right)=\bg\psi_{0}\left(\xi^{1},\xi^{2}\right)+\xi^{3}h_{0}\mathbf{G}\left(\xi^{1},\xi^{2}\right)=\bg\psi_{0}\left(\xi^{1},\xi^{2}\right)+\xi^{3}\mathbf{D}\left(\xi^{1},\xi^{2}\right)\label{eq1}
\end{equation}
where $\bg\psi_{0}$ is the reference position vector of the shell
middle surface, $h_{0}$ is the initial shell thickness, $\mathbf{G}$
is the initial unit normal, and $\mathbf{D}=h_{0}\mathbf{G}$, with
$\left\Vert \mathbf{G}\right\Vert =1$ and $\left\Vert \mathbf{D}\right\Vert =h_{0}$.

The position vector in the current shell configuration is defined
as :
\begin{equation}
\mathbf{x}\left(\xi^{1},\xi^{2},\xi^{3}\right)=\bg\psi\left(\xi^{1},\xi^{2}\right)+\xi^{3}h\left(\xi^{1},\xi^{2}\right)\mathbf{g}\left(\xi^{1},\xi^{2}\right)=\bg\psi\left(\xi^{1},\xi^{2}\right)+\xi^{3}\mathbf{d}\left(\xi^{1},\xi^{2}\right)\label{eq6}
\end{equation}
where $\bg\psi$ is the current position vector of the shell middle
surface, $h$ is the current shell thickness, $\mathbf{g}$ is the
current unit normal, and $\mathbf{d}=h\mathbf{g}$, with $\left\Vert \mathbf{g}\right\Vert =1$
and $\left\Vert \mathbf{d}\right\Vert =h$. In turn, $\bg\psi$ can
be expressed as
\begin{equation}
\bg\psi\left(\xi^{1},\xi^{2}\right)=\bg\psi_{0}\left(\xi^{1},\xi^{2}\right)+\mathbf{u}\left(\xi^{1},\xi^{2}\right)\label{eq7}
\end{equation}
with $\mathbf{u}$ as the shell middle surface displacement vector.
By introducing the thickness stretch
\begin{equation}
\lambda\left(\xi^{1},\xi^{2}\right)=\frac{h\left(\xi^{1},\xi^{2}\right)}{h_{0}}\label{eq9}
\end{equation}
the current shell director can be rewritten as

\begin{equation}
\mathbf{d}=h_{0}\lambda\mathbf{g}
\end{equation}

Let us now introduce the covariant basis vectors in the reference
and current configurations, respectively $\mathbf{G}_{i}$ and $\mathbf{g}_{i}$,
as follows, see Fig. \ref{fig:7}

\begin{equation}
\mathbf{G}_{\alpha}=\mathbf{X}_{,\alpha}=\bg\psi_{0,\alpha}\left(\xi^{1},\xi^{2}\right)+\xi^{3}\mathbf{D}_{,\alpha}\left(\xi^{1},\xi^{2}\right)\qquad\mathbf{G}_{3}=\mathbf{X}_{,3}=\mathbf{D}\label{eq:G}
\end{equation}

\begin{equation}
\mathbf{g}_{\alpha}=\mathbf{x}_{,\alpha}=\bg\psi_{,\alpha}\left(\xi^{1},\xi^{2}\right)+\xi^{3}\mathbf{d}_{,\alpha}\left(\xi^{1},\xi^{2}\right)\qquad\mathbf{g}_{3}=\mathbf{x}_{,3}=\mathbf{d}\label{eq:g}
\end{equation}
where

\begin{equation}
\mathbf{D}_{,\alpha}=h_{0}\mathbf{G}_{,\alpha}\quad\mathbf{d}_{,\alpha}=h_{0}\left(\lambda_{,\alpha}\mathbf{g}+\lambda\mathbf{g}_{,\alpha}\right)
\end{equation}
From the covariant bases, the contravariant bases $\mathbf{G}^{i}$
and $\mathbf{g}^{i}$ are induced with
\begin{equation}
\mathbf{G}_{i}\cdot\mathbf{G}^{j}=\mathbf{g}_{i}\cdot\mathbf{g}^{j}=\delta_{i}^{j}=\left\{ \begin{array}{cc}
0 & i\neq j\\
1 & i=j
\end{array}\right.\label{eq14}
\end{equation}

and

\begin{equation}
G_{ij}=\mathbf{G}_{i}\cdot\mathbf{G}_{j}\quad g_{ij}=\mathbf{g}_{i}\cdot\mathbf{g}_{j}\label{eq:gij}
\end{equation}

The deformation gradient $\mathbf{F}=\partial\mathbf{x}/\partial\mathbf{X}$
can be expressed as a function of the basis vectors in the reference
and current configurations:

\begin{equation}
\mathbf{F}=\mathbf{g}_{i}\otimes\mathbf{G}^{i}\quad\mathbf{F}^{T}=\mathbf{G}^{i}\otimes\mathbf{g}_{i}\quad\mathbf{F}^{-1}=\mathbf{G}_{i}\otimes\mathbf{g}^{i}\quad\mathbf{F}^{-T}=\mathbf{g}^{i}\otimes\mathbf{G}_{i}\label{eq:F}
\end{equation}

A suitable strain measure is the Green-Lagrange strain tensor, defined
as

\begin{equation}
\mathbf{E}=\frac{1}{2}\left(\mathbf{F}^{T}\mathbf{F}-\mathbf{1}\right)=E_{ij}\mathbf{G}^{i}\otimes\mathbf{G}^{j}\label{eq:E}
\end{equation}
where $\mathbf{1}$ is the second-order unit tensor. Substituting
eq. (\ref{eq:F}) into eq. (\ref{eq:E}) and recalling that the identity
tensor is identical to the metric tensor yields

\begin{equation}
\mathbf{E}=\frac{1}{2}\left[\left(\mathbf{G}^{i}\otimes\mathbf{g}_{i}\right)\cdot\left(\mathbf{g}_{j}\otimes\mathbf{G}^{j}\right)-G_{ij}\mathbf{G}^{i}\otimes\mathbf{G}^{j}\right]=\frac{1}{2}\left(g_{ij}-G_{ij}\right)\mathbf{G}^{i}\otimes\mathbf{G}^{j}\label{eq:E_bis}
\end{equation}
Comparison of eqs. (\ref{eq:E}) and (\ref{eq:E_bis}) yields

\begin{equation}
E_{ij}=\frac{1}{2}\left(g_{ij}-G_{ij}\right)\label{eq:Eij}
\end{equation}

If the expressions of $g_{\text{ij}}$ and $G_{\text{ij}}$ in eq.
(\ref{eq:Eij}) are expanded using eqs. (\ref{eq:gij}), (\ref{eq:G})
and (\ref{eq:g}), the in-plane strain components can be expressed
as follows:
\begin{equation}
E_{\alpha\beta}=\epsilon_{\alpha\beta}+\xi^{3}\kappa_{\alpha\beta}\label{eq: Eab}
\end{equation}
with
\begin{equation}
\epsilon_{\alpha\beta}=\frac{1}{2}\left(\bg\psi_{,\alpha}\cdot\bg\psi_{,\beta}-\bg\psi_{0,\alpha}\cdot\bg\psi_{0,\beta}\right)\label{eq24}
\end{equation}
representing the membrane strain components and
\begin{equation}
\kappa_{\alpha\beta}=\frac{1}{2}\left(\bg\psi_{,\alpha}\cdot\mathbf{d}_{,\beta}+\bg\psi_{,\beta}\cdot\mathbf{d}_{,\alpha}-\bg\psi_{0,\alpha}\cdot\mathbf{D}{}_{,\beta}-\bg\psi_{0,\beta}\cdot\mathbf{D}{}_{,\alpha}\right)\label{eq25}
\end{equation}
representing the change of curvature. In eq. (\ref{eq: Eab}) an additional
term quadratic in $\xi^{3}$ has been neglected, \color{black} a usual assumption which delivers accurate results for thin shells such as those considered herein \cite{key-901,key-902}. \color{black} The shear strain
components assume the form
\begin{equation}
\gamma_{\alpha}=2E_{\text{\ensuremath{\alpha}3}}\label{eq27}
\end{equation}
where
\begin{equation}
E_{\text{\ensuremath{\alpha}3}}=\frac{1}{2}\left(\bg\psi_{,\alpha}\cdot\mathbf{d}-\bg\psi_{0,\alpha}\cdot\mathbf{D}\right)\label{eq28}
\end{equation}
again neglecting an additional term, linear in $\xi^{3}$. Finally,
the thickness strain is obtained as
\begin{equation}
E_{33}=\frac{1}{2}\left(\left\Vert \mathbf{d}\right\Vert ^{2}-\left\Vert \mathbf{D}\right\Vert ^{2}\right)=\frac{1}{2}\left(\left\Vert \mathbf{d}\right\Vert ^{2}-h_{0}^{2}\right)\label{eq29}
\end{equation}

\subsection{{\normalsize{}Electric field}}

The electric field is given by the gradient of the electric potential
$\phi$, hence in the material configuration its expression is given by
\color{black}
\begin{equation}
\vec{\textbf{E}}=-\phi_{,i} \mathbf{G}^{i}
\end{equation}
\color{black}
If the piezoelectric material is assumed to be poled in the shell
thickness direction, only the difference of electric potential in
this direction must be considered, otherwise also the other contributions
are required.
\color{black}
A common assumption in piezoelectric models is that
the electric field is constant through the thickness inside the actuator or sensor. This is in bending dominated situations not correct [27]. Herein, a linear approximation is adopted, which is sufficient to pass the out-of-plane bending patch test [27].
\color{black}

\subsection{{\normalsize{}Generalized strain vector}}

The Green-Lagrange strain and the electric field components of the
piezoelectric solid can be arranged in a generalized strain column
vector $\mathbf{E}_{g}$:
\begin{equation}
\mathbf{E}_{g}=\left[E_{11},E_{22},E_{33},2E_{12},2E_{13},2E_{23},\vec{E}_{1},\vec{E}_{2},\vec{E}_{3}\right]^{T}\label{eq34}
\end{equation}
whereas the strain and electric field components of the piezoelectric
shell can be arranged in the following vector
\begin{equation}
\mathbf{E}_{s}=\left[\epsilon_{11},\epsilon_{22},2\epsilon_{12},\kappa_{11},\kappa_{22},2\kappa_{12},\gamma_{1},\gamma_{2},\vec{E}_{1},\vec{E}_{2},\epsilon^{0}{}_{33},\epsilon^{1}{}_{33},\vec{E}^{0}{}_{3},\vec{E}^{1}{}_{3}\right]^{T}\label{eq35}
\end{equation}
where ${\epsilon}^{0}{}_{33},{\epsilon}^{1}{}_{33}$ are the constant
and linear components of the thickness strain, while ${\vec{E}}^{0}{}_{3},{\vec{E}}^{1}{}_{3}$
represent the constant and linear parts of the electric field in the
thickness direction \color{black}\cite{key-14}\color{black}. The relation between the Green-Lagrange strains
and the independent shell strains can be stated in compact form by
introducing the matrix $\textbf{A}$ such as:
\begin{equation}
\mathbf{E}_{g}=\mathbb{\mathbf{A}}\mathbf{E}_{s}\label{eq36}
\end{equation}
where:
\begin{equation}
\mathbf{A}=\left(\begin{array}{cccccccccccccc}
1 & 0 & 0 & \xi^{3} & 0 & 0 & 0 & 0 & 0 & 0 & 0 & 0 & 0 & 0\\
0 & 1 & 0 & 0 & \xi^{3} & 0 & 0 & 0 & 0 & 0 & 0 & 0 & 0 & 0\\
0 & 0 & 0 & 0 & 0 & 0 & 0 & 0 & 0 & 0 & 1 & \xi^{3} & 0 & 0\\
0 & 0 & 1 & 0 & 0 & \xi^{3} & 0 & 0 & 0 & 0 & 0 & 0 & 0 & 0\\
0 & 0 & 0 & 0 & 0 & 0 & 1 & 0 & 0 & 0 & 0 & 0 & 0 & 0\\
0 & 0 & 0 & 0 & 0 & 0 & 0 & 1 & 0 & 0 & 0 & 0 & 0 & 0\\
0 & 0 & 0 & 0 & 0 & 0 & 0 & 0 & 1 & 0 & 0 & 0 & 0 & 0\\
0 & 0 & 0 & 0 & 0 & 0 & 0 & 0 & 0 & 1 & 0 & 0 & 0 & 0\\
0 & 0 & 0 & 0 & 0 & 0 & 0 & 0 & 0 & 0 & 0 & 0 & 1 & \xi^{3}
\end{array}\right)\label{eq37}
\end{equation}

\section{{\normalsize{} \color{black} The microscale boundary value problem}}
\color{black}
The material within the RVE consists in piezoelectric polymer fibers
with a tight packing arrangement (Fig. \ref{fig:6}). The perfectly
aligned configuration considered herein is a simplification of the
actual geometry, which will be tackled in future extensions. The fibers are partially
bonded to each other (as will be described \color{black} in more detail \color{black} in Section
\ref{sub:Boundary-conditions}) and partially subjected to unilateral
electromechanical contact constraints.
This section overviews \color{black}the kinematically nonlinear electroelasticity
theory \color{black} and the contact formulation, whereas
the boundary conditions will be discussed in Section \ref{sub:Boundary-conditions}
as they are integral to the scale transition procedure.

\color{black}
\subsection{{\normalsize{} Nonlinear electro-elasticity theory}}
The starting point for the variational setting of geometrically nonlinear, quasi-static electro-elasticity is the definition of an energy density per
\color{black} unit reference volume \cite{key-1177}:
\begin{equation}\label{1}
H^m(\textbf{F}^m, \vec {\textbf{E}} ^m)
\end{equation}
where $\textbf{F}^m$ is the deformation gradient and $\vec {\textbf{E}} ^m$ is the electric field in the reference configuration given by $\vec {\textbf{E}} ^m=-Grad\phi^m$ where $Grad$ is the gradient operator in the reference configuration and $\phi^m$ is the electric potential. All quantities refer to the microscale as indicated by the superscript m, also used in the following for the same purpose. Based on the above energy density the first Piola Kirchoff stress $\textbf{P}^m$ and the reference dielectric displacement $\vec {\textbf{D}} ^m$ are defined as:
\begin{equation}\label{2}
\textbf{P}^m = \frac {\partial H^m} {\partial \textbf{F}^m}; \vec {\textbf{D}} ^m=-\frac{\partial H^m}{\partial \vec {\textbf{E}} ^m}
\end{equation}
 The mechanical equilibrium conditions and the Gauss law of electricity are (in the absence of body forces and charges):
\begin{equation}\label{3}
Div\textbf{P}^m=\textbf{0}; Div\vec {\textbf{D}} ^m=0  ;
\end{equation}
where $Div$ is the divergence operator.
The fourth-order reference elasticity tensor $\pmb{\mathfrak{C}}^m$ follows as the derivative of the first total Piola Kirchoff stress with respect to the deformation gradient $\textbf{F}^m$ or as part of the hessian of the total energy density $H^m$:
\begin{equation}\label{eq:e}
\pmb{\mathfrak{C}}^m=\frac{\partial \textbf{P}^m}{\partial \textbf{F}^m}=\frac{\partial ^2H^m}{\partial \textbf{F}^m \otimes \partial \textbf{F}^m}
\end{equation}
Likewise the second-order reference dielectricity tensor $\tilde{\pmb{\mathfrak{e}}}^m$ follows as the derivative of the reference dielectric displacement with respect to the reference electric field or as part of the hessian of the total energy density $H^m$:
\begin{equation}\label{eq:e}
\tilde{\pmb{\mathfrak{e}}}^m=\frac{\partial \vec{\textbf{D}}^m}{\partial \vec{\textbf{E}}^m}=-\frac{\partial ^2H^m}{\partial \vec{\textbf{E}}^m \otimes \partial \vec{\textbf{E}}^m}
\end{equation}
Finally the third-order reference piezoelectricity tensor $\pmb{\mathfrak {E}}^m$ follows either as the derivative of the reference dielectric displacement with respect to the deformation gradient or alternatively as the derivative of the first Piola Kirchoff stress with respect to the reference electric field:
\begin{center}
\begin{equation}\label{6}
\pmb{\mathfrak {E}}^m=\frac{\partial \vec{\textbf{D}}^m}{\partial \textbf{F}^m}=-(\frac{\partial \textbf{P}^m}{\partial \vec{\textbf{E}}^m})^T=-\frac{\partial ^2H^m}{\partial \vec{\textbf{E}}^m \otimes \partial \textbf{F}^m}
\end{equation}
\end{center}

Objectivity requires that (with a slight notation abuse):
\begin{equation}\label{obj}
 H^m(\textbf{F}^m, \vec {\textbf{E}} ^m)=H^m(\textbf{C}^m, \vec {\textbf{E}} ^m)
\end{equation}
where $\textbf{C}^m=(\textbf{F}^m)^T\textbf{F}^m$ is the right Cauchy Green tensor, or equivalently:
\begin{equation}\label{obj}
 H^m(\textbf{F}^m, \vec {\textbf{E}} ^m)=H^m(\textbf{E}^m, \vec {\textbf{E}} ^m)
\end{equation}
where $\textbf{E}^m=\frac{1}{2} \left(\textbf{C}^m -\textbf{1}\right)$ is the Green Lagrange strain tensor.
Moreover instead of $\textbf{P}^m$ it is often more convenient to use the second
Piola Kirchhoff  stress $\textbf{S}^m=(\textbf{F}^{m})^{-1} \textbf{P}^m$. So we can now introduce the fourth-order reference elasticity tensor and the third-order reference piezoelectricity tensor referred to the
new variables $\textbf{S}^m$ and $\textbf{E}^m$ as

\begin{equation}\label{eq:c2}
\pmb{\mathfrak{c}}^m=\frac{\partial \textbf{S}^m}{\partial \textbf{E}^m}=\frac{\partial ^2H^m}{\partial \textbf{E}^m \otimes \partial \textbf{E}^m}
\end{equation}

\begin{center}
\begin{equation}\label{eq:e2}
\pmb{\mathfrak {e}}^m=\frac{\partial \vec{\textbf{D}}^m}{\partial \textbf{E}^m}=-\frac{\partial ^2H^m}{\partial \vec{\textbf{E}}^m \otimes \partial \textbf{E}^m}
\end{equation}
\end{center}

All constants terms can be grouped in a generalized microscale constitutive matrix $\mathbb{\mathbf{D}}_{\text{}}^{m}$:
\begin{equation}
\mathbb{\mathbf{D}}_{\text{}}^{m}=
\left[\begin{array}{cc}
\pmb{\mathfrak{c}}^m &
-\pmb{\mathfrak{e}}^{m^T}\\
\pmb{\mathfrak{e}}^m &
\tilde{\pmb{\mathfrak{e}}}^m
\end{array}\right]
\end{equation}
where $\pmb{\mathfrak{c}}^m$ and $\pmb{\mathfrak{e}}^m$ are matrices obtained respectively
from the fourth and third order tensors in eqs. (\ref{eq:c2}) and
(\ref{eq:e2}) using Voigt notation. Further assuming linear behavior, we can write the piezoelectric constitutive equation as:
\begin{equation}
\begin{bmatrix}S_{11}^m\\
S_{22}^m\\
S_{33}^m\\
S_{12}^m\\
S_{13}^m\\
S_{23}^m\\
\vec{D}_{1}^m\\
\vec{D}_{2}^m\\
\vec{D}_{3}^m
\end{bmatrix}=\mathbf{D}^{m}\begin{bmatrix}E_{11}^m\\
E_{22}^m\\
E_{33}^m\\
2E_{12}^m\\
2E_{13}^m\\
2E_{23}^m\\
\vec{E}_{1}^m\\
\vec{E}_{2}^m\\
\vec{E}_{3}^m
\end{bmatrix};
\end{equation}
The Dirichlet and the Neumann boundary conditions for the mechanical field, see Fig. \ref{fig:7b}, are
\begin{equation}
\textbf{u}^{m}=\bar{\textbf{u}}^{m}\text{ }\text{on }\Gamma_{u}\quad \textbf{t}^m=\textbf{P}^m \cdot \textbf{N}=\bar{\textbf{t}}^m\text{ on }\Gamma_{t}\label{eq45}
\end{equation}
where $\bar{\mathbf{u}}^m$ and $\bar{\mathbf{t}}^m$ are prescribed mechanical
displacement and surface traction vectors in the reference configuration, and $\Gamma=\Gamma_{u}\cup\Gamma_{t}$,
$\Gamma_{u}\cap\Gamma_{t}=\oslash$, with $\Gamma$ as the boundary
of the domain and $\Gamma_{u}$, $\Gamma_{t}$ as its Dirichlet and Neumann portions. Moreover $\textbf{N}$ is the outward unit normal to $\Gamma$. The boundary conditions for the electric field are
\begin{equation}
\phi^m=\bar{\phi}^m\text{ }\text{on }\Gamma_{\phi}\quad\vec{d}^m=\vec{\textbf{D}}^m \cdot \textbf{N}=\vec{\bar{d}}\text{ on }\Gamma_{\vec{\, d}}\label{eq46}
\end{equation}
where $\bar{\phi}^m$ and $\vec{\bar{d}}^m$ are prescribed values of
electric potential and electric charge flux, and $\Gamma=\Gamma_{\phi}\cup\Gamma_{\vec{d}}$,
$\Gamma_{\phi}\cap\Gamma_{\vec{d}}=\oslash$.

\subsection{{\normalsize{}Finite element formulation including electromechanical
contact}}
Within the RVE, the piezoelectric fibers are subjected to contact
constraints. \color{black} For the sake of simplicity,
we assume herein frictionless contact conditions. Since no experimental results are available on contact interactions between the piezoelectric fibers examined in this paper, the choice of a friction law and of the related coefficient(s) would have been arbitrary. The addition of frictional effects and the corresponding experimental justification may well be considered in further developments of this research. \color{black}
Herein the main characteristics of the 3D electromechanical frictionless contact formulation are now summarized.
All contact related quantities are referred to the micro scale. The formulation is
based on the classical master-slave concept \cite{key-100}, see Fig.
\ref{fig:7c}. For each point on the slave surface, $\mathbf{x}^{s}$,
the corresponding point on the master surface is determined through
normal (i.e. closest point) projection and is denoted as $\bar{\mathbf{x}}^{\mathfrak {m}}$.
Thus the normal gap of each slave point is computed as:
\begin{equation}
g_{N}=\left(\mathbf{x}^{s}-\bar{\mathbf{x}}^{\mathfrak {m}}\right)\cdot\mathbf{\bar{n}}\label{eq: gap}
\end{equation}
$\mathbf{\bar{n}}$ being the outer unit normal to the master surface
at the projection point \color{black} in the current configuration\color{black}. The sign of the measured gap is used to discriminate
between active and inactive contact conditions, a negative value of
the gap leading to active contact. The electric field requires the
definition of the contact electric potential jump, $g_{\phi}$:
\begin{equation}
g_{\phi}=\phi^{s}-\bar{\phi}^{\mathfrak {m}}\label{eq: el_gap}
\end{equation}
where $\phi^{s}$ and $\bar{\phi}^{\mathfrak {m}}$ are the electric potential
values in the slave point and in its master projection point.

The discretization strategy used herein for the contact contribution
is based on the node-to-surface approach combined with Bézier smoothing
of the master surface. It is well known that the node-to-surface algorithm
is susceptible of pathologies due to the $C^{0}$-continuity of the
finite element (Lagrange) discretizations, and that these may affect
the quality of results as well as iterative convergence in contact
computations \cite{key-23,key-101}. One of the possible remedies are smoothing
techniques for the master surface. Herein, the technique based on
Bézier patches proposed by \cite{key-21} and implemented within the
AceGen/AceFEM environment is adopted and straightforwardly extended
to electromechanical contact constraints, see Fig. \ref{fig:7c}.
Here a tensor product representation of one-dimensional Bézier polynomials
is used to interpolate the master surfaces in the contact interface.
A three-dimensional Bézier surface of order $\mathfrak{N}$ is given by
\begin{equation}
\mathbf{x}(\zeta^{\alpha})=\sum_{k=0}^{\mathfrak {N}}\sum_{l=0}^{\mathfrak {N}}B_{k}^{\mathfrak {N}}(\zeta^{1})B_{l}^{\mathfrak {N}}(\zeta^{2})\mathbf{d}_{kl}\label{eq47}
\end{equation}
where $B_{k}^{\mathfrak {N}}(\zeta^{1})$ and $B_{l}^{\mathfrak {N}}(\zeta^{2})$ are Bernstein polynomials, functions of the convective surface coordinates $\zeta^{\alpha}$, and $\mathbf{d}_{kl}$ are the coordinates of the so-called control points. To fully characterize a bicubic Bézier patch, 16 control points $\mathbf{d}_{kl}$ are needed.

The technique used herein consists in the so-called Bézier-9 patches
\cite{key-21}. Each patch is constructed using one central node of
the master surface, denoted as $\mathbf{x}_{22}^{\mathfrak {m}}$, and eight neighbouring
nodes $\mathbf{x}_{ij}^{\mathfrak {m}}$, $i,j=1,...,3$ ($\left\{ i,j\right\} \neq\left\{ 2,2\right\} $)
forming four bilinear segments surrounding the central node. The 16
control points needed to define the bicubic Bézier patch are obtained
from

\begin{eqnarray}
\mathbf{d}_{\text{k1}}=\frac{1}{2}\left(\hat{\mathbf{d}}_{\text{k1}}+\hat{\mathbf{d}}_{\text{k2}}\right) &  & \mathbf{d}_{\text{k2}}=\frac{1}{2}(1-\beta)\left(\hat{\mathbf{d}}_{\text{k2}}+\hat{\mathbf{d}}_{\text{k1}}\right)+\beta\hat{\mathbf{d}}_{\text{k2}}\label{eq49}\\
\mathbf{d}_{\text{k3}}=\frac{1}{2}(1-\beta)\left(\hat{\mathbf{d}}_{\text{k2}}+\hat{\mathbf{d}}_{\text{k3}}\right)+\beta\hat{\mathbf{d}}_{\text{k2}} &  & \mathbf{d}_{\text{k4}}=\frac{1}{2}\left(\hat{\mathbf{d}}_{\text{k3}}+\hat{\mathbf{d}}_{\text{k2}}\right)
\end{eqnarray}
where $k=1,...,4$, $\beta$ is a parameter defining the shape of
the surface (here $\beta=2/3$), and the auxiliary points $\hat{\mathbf{d}}_{\text{kj}}$
are defined in terms of the master nodes $\mathbf{x}_{\text{ij}}^{\mathfrak {m}}$ according to
\begin{eqnarray}
\hat{\mathbf{d}}_{1j}=\frac{1}{2}(\mathbf{x}_{1j}^{\mathfrak {m}}+\mathbf{x}_{2j}^{\mathfrak {m}}) &  & \hat{\mathbf{d}}_{2j}=\frac{1}{2}(1-\beta)(\mathbf{x}_{2j}^{\mathfrak {m}}+\mathbf{x}_{ij}^{\mathfrak {m}})+\beta\mathbf{x}_{2j}^{\mathfrak {m}}\label{eq50}\\
\hat{\mathbf{d}}_{3j}=\frac{1}{2}(1-\beta)(\mathbf{x}_{2j}^{\mathfrak {m}}+\mathbf{x}_{3j}^{\mathfrak {m}})+\beta\mathbf{x}_{2j}^{\mathfrak {m}} &  & \hat{\mathbf{d}}_{4j}=\frac{1}{2}(\mathbf{x}_{3j}^{\mathfrak {m}}+\mathbf{x}_{2j}^{\mathfrak {m}})
\end{eqnarray}
The same procedure is followed for the electric potential, which is
expressed as
\begin{equation}
\phi(\zeta^{\alpha})=\sum_{k=0}^{\mathfrak {
N}}\sum_{l=0}^{\mathfrak {N}}B_{k}^{\mathfrak {N}}(\zeta^{1})B_{l}^{\mathfrak {N}}(\zeta^{2})\phi_{\text{kl}}\label{eq51}
\end{equation}
where now $\phi_{\text{kl}}$ is the potential evaluated at the 16
control points as a function of the potential at the auxiliary points
$\hat{\phi}_{\text{kj}}$ and at the master nodes $\phi_{\text{ij}}^{\mathfrak {m}}$
.

Note that the need for smoothing of the master surface can be straightforwardly
eliminated by the use of isogeometric discretizations, as these feature
higher continuity and smoothness than classical Lagrange discretizations
(see e.g. \cite{key-24}). This will be done in
future extensions of the present work.

According to standard finite element techniques, the global set of
equations can be obtained by adding to the variation of the energy
potential representing the continuum behavior the virtual work due
to the electromechanical contact contribution associated to the active
contact elements. The global energy of the discretized system $\Pi^{m}$
(where $m$ refers again to the microscale) is thus
\color{black}
\begin{equation}
\Pi^{m}=\cup H^m(\textbf{E}^m, \vec {\textbf{E}} ^m)+\underset{\text{active}}{\cup}\Pi_{C}^{m}\label{eq55}
\end{equation}
\color{black}
The virtual work due to contact is in turn given by
\begin{equation}
\Pi_{C}^{m}=\Pi_{MC}^{m}+\Pi_{EC}^{m}
\end{equation}
with $\Pi_{MC}^{m}$ and $\Pi_{EC}^{m}$ as the virtual work due to
mechanical and electric contact, respectively. Herein, the electromechanical
contact constraints are regularized with the penalty method, which
leads to
\begin{equation}
\Pi_{MC}^{m}=\frac{1}{2}F_{N}g_{N}\quad\Pi_{EC}^{m}=\frac{1}{2}I_{N}g_{\phi}
\end{equation}
where
\begin{equation}
F_{N}=\rho_{\text{mech}}g_{N}\quad I_{N}=\rho_{\text{el}}g_{\phi}\label{54}
\end{equation}
are respectively the contact force and the electric current, and $\rho_{\text{mech}}$
and $\rho_{\text{el}}$ are penalty parameters. Virtual variation
of eq. (\ref{eq55}) gives
\color{black}
\begin{equation}
\delta_{u}\left(\Pi^{m}\right)=\cup\delta_{u}(H^m(\textbf{E}^m, \vec {\textbf{E}} ^m))+\underset{\text{active}}{\cup}\left[\rho_{\text{mech}}g_{N}\delta_{u}g_{N}\right]\label{eq52}
\end{equation}
\begin{equation}
\delta_{\phi}\left(\Pi^{m}\right)=\cup\delta_{\phi}(H^m(\textbf{E}^m, \vec {\textbf{E}} ^m))+\underset{\text{active}}{\cup}\left[\rho_{\text{el}}g_{\phi}\delta_{\phi}g_{\phi}\right]\label{eq53}
\end{equation}
\color{black}
and its consistent linearization leads to
\color{black}
\begin{align}
\Delta_{u}\delta_{u}\left(\Pi^{m}\right)=\cup\Delta_{u}\delta_{u}(H^m(\textbf{E}^m, \vec {\textbf{E}} ^m))+\underset{\text{active}}{\cup}\Delta_{u}\left[\rho_{\text{mech}}g_{N}\delta_{u}g_{N}\right]\\
\Delta_{\phi}\delta_{u}\left(\Pi^{m}\right)=\cup\Delta_{\phi}\delta_{u}(H^m(\textbf{E}^m, \vec {\textbf{E}}^m))+\underset{\text{active}}{\cup}\Delta_{\phi}\left[\rho_{\text{mech}}g_{N}\delta_{u}g_{N}\right]\\
\Delta_{u}\delta_{\phi}\left(\Pi^{m}\right)=\cup\Delta_{u}\delta_{\phi}(H^m(\textbf{E}^m, \vec {\textbf{E}}^m))+\underset{\text{active}}{\cup}\Delta_{u}\left[\rho_{\text{el}}g_{\phi}\delta_{\phi}g_{\phi}\right]\\
\Delta_{\phi}\delta_{\phi}\left(\Pi^{m}\right)=\cup\Delta_{\phi}\delta_{\phi}(H^m(\textbf{E}^m, \vec {\textbf{E}} ^m))+\underset{\text{active}}{\cup}\Delta_{\phi}\left[\rho_{\text{el}}g_{\phi}\delta_{\phi}g_{\phi}\right]
\end{align}

The advanced symbolic computational tools available in the AceGen/AceFEM
finite element environment allow for a full automation of the linearization
process, see \cite{key-20} for more details. {\small{}If $\mathbf{u}^m=N_{i}\hat{\mathbf{u}}_{i}^m$
and $\phi^m=N_{i}\hat{\phi_{i}}^m$ }are the discretized displacement
and electric potential fields,{\small{} }with{\small{} $\hat{\mathbf{u}}_{i}^m$
and $\hat{\phi_{i}}^m$ }as the nodal displacements and electric potential,
respectively, and{\small{} $N_{i}$ }as the (in this case linear)
shape functions, the residual vector and the stiffness matrix terms
resulting from the finite element discretization are determined as
follows{\small{}
\begin{equation}
\mathbf{R}_{u_{i}}^m=\frac{\delta\Pi^{m}}{\delta\hat{\mathbf{u}}{}_{i}^m}\quad R_{\phi_{i}}^m=\frac{\delta\Pi^{m}}{\delta\hat{\phi}_{i}^m}
\end{equation}

\begin{align}
\mathbf{K}_{\text{uu}_{ij}}^m=\frac{\delta R_{u_{i}}^m}{\delta\hat{\mathbf{u}}{}_{j}^m}\quad & K_{\phi\phi_{ij}}^m=\frac{\delta R_{\phi_{i}}^m}{\delta\hat{\phi}{}_{j}^m}\\
\mathbf{K}_{\text{u\ensuremath{\phi}}_{ij}}^m=\frac{\delta\mathbf{R}_{u_{i}}^m}{\delta\hat{\phi}{}_{j}^m}\quad & \mathbf{K}_{\text{\ensuremath{\phi}u}_{ij}}^m=\frac{\delta R_{\phi_{i}}^m}{\delta\hat{\mathbf{u}}{}_{j}^m}
\end{align}
}{\small \par}
\color{black}
\section{{\normalsize{}Computational homogenization procedure}}

In this section, a two-step homogenization method is introduced. Starting
from the microscale RVE, we obtain in the first step the macroscopic
constitutive response of an equivalent homogenized solid, and in the
second step, through thickness integration, the effective coefficients
of an homogenized shell. First, the concept of RVE and Hill's energy
principle are briefly introduced. This allows for the transition from
micro to macro quantities characterizing the multiphysics problem.
Then, a method to calculate the macroscopic constitutive matrix starting
from the solution of a microscale boundary value problem (BVP) is
described for the general case of electromechanical problems.

\subsection{{\normalsize{}Unit cell models for numerical homogenization}}
\color{black}
The main idea of homogenization is finding an homogeneous medium equivalent
to the original heterogeneous material, such that the strain energies
stored in the two systems are the same (Hill condition). Split of
the microscopic deformation gradient and electric field into constant parts $\bar{\textbf{F}}^m,\bar{\vec{\textbf{E}}}^{m}$
and fluctuating parts $\nabla\tilde{\textbf{u}}^{m},\tilde{\vec{\textbf{E}}}^{m}$
yields:
\begin{equation}
\textbf{F}^m=\bar{\textbf{F}}^m+\nabla\tilde{\textbf{u}}^{m};\text{ }
\vec{\textbf{E}}^{m}=\bar{\vec{\textbf{E}}}^{m}+\tilde{\vec{\textbf{E}}}^{m}
\end{equation}
where by definition
\begin{equation}
\int_{V}\nabla\tilde{\textbf{u}}^{m}dV=0\text{ }\quad\text{ }\int_{V}\tilde{\vec{\textbf{E}}}^{m}dV=0
\end{equation}
In the previous and in the next equations (unless otherwise specified),
the RVE volume in the reference configuration $V_{RVE}$, Fig. \ref{fig:8}, is indicated
as $V$ for the sake of a compact notation. Formulated for the electromechanical
problem at hand, the Hill criterion in differential form reads
\begin{equation}
\bar{\textbf{P}}^{M}:\delta\bar{\textbf{F}}^{M}+\bar{\vec{\textbf{D}}}^{M} \cdot \delta\bar{\vec{\textbf{E}}}^{M}
=\frac{1}{V}\int_{V}\textbf{P}^{m}:\delta\textbf{F}^{m}dV+\frac{1}{V}\int_{V}\vec{\textbf{D}}^{m} \cdot \delta\vec{\textbf{E}}^{m}dV\label{eq: Hill}
\end{equation}
and requires that the macroscopic volume average of the variation
of work performed on the RVE is equal to the local variation of the
work on the macroscale. In the previous equation as well as in the following,
the superscript M refers to the macroscale. In particular, $\bar{\textbf{P}}^{M}$,
$\bar{\vec{\textbf{D}}}^{M}$, $\bar{\textbf{F}}^{M}$, and $\bar{\vec{\textbf{E}}}^{M}$
represent respectively the average values of the first Piola Kirchhoff stress, electric displacement, deformation gradient and electric field, i.e.

\begin{equation}
\bar{\textbf{F}}^M=\frac{1}{V}\int_{V}\textbf{F}^{m}dV\quad\bar{\textbf{P}}^{M}=\frac{1}{V}\int_{V}\textbf{P}^{m}dV\label{eq:s_av}
\end{equation}
\begin{equation}
\bar{\vec{\textbf{E}}}^{M}=\frac{1}{V}\int_{V}\vec{\textbf{E}}^{m}dV\quad\bar{\vec{\textbf{D}}}^{M}=\frac{1}{V}\int_{V}\vec{\textbf{D}}^{m}dV\label{eq:e_av}
\end{equation}
Eq. \ref{eq: Hill} may be split into two parts

\begin{equation}
\bar{\textbf{P}}^{M}:\delta\bar{\textbf{F}}^{M}=\frac{1}{V}\int_{V}\textbf{P}^{m}:\delta\textbf{F}^{m}dV;
\quad\bar{\vec{\textbf{D}}}^{M} \cdot \delta\bar{\vec{\textbf{E}}}^{M}=\frac{1}{V}\int_{V}\vec{\textbf{D}}^{m} \cdot \delta\vec{\textbf{E}}^{m}dV
\label{eq: Hill2}
\end{equation}
\color{black}
\subsection{{\normalsize{}\label{sub:Boundary-conditions}Boundary conditions}}

Classically three types of boundary conditions are used for an RVE
\cite{key-13,key-27}: prescribed linear displacements, prescribed
constant tractions and periodic boundary conditions. All three types
of boundary conditions satisfy the micro-macro work equality stemming
from Hill's lemma in eq. (\ref{eq: Hill}) and are therefore suitable
for the analysis. Periodic boundary conditions (Fig. \ref{fig:9})
are suitable for solids with periodic microstructure (Fig. \ref{fig:6}),
such as the PVDF sheet under study, and therefore are preferred in
this paper. The periodicity conditions for the RVE are derived in
a general format as: \color{black}
\begin{align}
u_{i}^{K^{+}}=u_{i}^{K^{-}}\text{ }\quad t_{i}^{K^{+}}=-t_{i}^{K^{-}}\\
\phi_{i}^{K^{+}}=\phi_{i}^{K^{-}}\text{ }\quad\vec{d}_{i}^{K^{+}}=-\vec{d}_{i}^{K^{-}}
\end{align} \color{black}
where the indices $K^{+}$ and $K^{-}$ indicate the values on two
opposite surfaces of the unit cell, and $u_{i}^{K^{+}}$, $u_{i}^{K^{-}}$,
$t_{i}^{K^{+}}$, $t_{i}^{K^{-}}$, $\phi_{i}^{K^{+}}$, $\phi_{i}^{K^{-}}$,
$\vec{d}_{i}^{K^{+}}$and $\vec{d}_{i}^{K^{-}}$, are respectively
the components of displacement, traction, electric potential, and
electric flux at corresponding points on the opposing faces $K^{+}$
and $K^{-}$ (Fig. \ref{fig:9}). Periodic boundary conditions are
implemented in Acegen as multipoint constraints using the Lagrange
multiplier method. Uniform meshing of the RVE facilitates the enforcement
of these constraints, as opposing nodes match on all the faces of
the RVE.
\color{black}
Note that the RVE structure in this study may be considered as a particular
case of bimaterial RVE, where the first material corresponds to the
fibers ($B$) and the second material is represented by the void
between the fibers in contact ($H$). In Fig. \ref{fig:8}, the
total RVE volume is thus $V_{RVE}=V_{B}\cup V_{H}$. During
volume integration the void part affects the average values of the
electromechanical properties.
\color{black}
Perfect bond between the fibers is assumed along the generatrices
of the cylinders (blue lines in Fig. \ref{fig:8}) and is enforced
through condensation of the respective degrees of freedom. This assumption
is based on observations of the fibers with the Scanning Electron
Microscope, which reveal as ``welding'' takes place between the
fibers upon manufacturing due to solvent evaporation \cite{key-2}.

\subsection{{\normalsize{}Macroscale constitutive law}}
\color{black}
Due to the electromechanical contact between fibers at
the microscale, the macroscopic constitutive law is nonlinear
but can be linearized introducing a tangent or secant constitutive matrix
for the macroproblem. In this work, we opted for the secant approach,
see also \cite{key-905,key-906,key-907}, in which the effective (or volume-averaged) generalized strain column vector of the homogenized material $\textbf{E}_g$ is computed successively (and independently) for increasing values of the effective generalized stress vector $\textbf{S}_g$. They are related in each step through a generalized secant effective constitutive matrix $\mathbf{D}^{M}$ so that
\begin{equation}
\begin{bmatrix}S_{11}\\
S_{22}\\
S_{33}\\
S_{12}\\
S_{13}\\
S_{23}\\
\vec{D}_{1}\\
\vec{D}_{2}\\
\vec{D}_{3}
\end{bmatrix}=\mathbf{D}^M\begin{bmatrix}E_{11}\\
E_{22}\\
E_{33}\\
2E_{12}\\
2E_{13}\\
2E_{23}\\
\vec{E}_{1}\\
\vec{E}_{2}\\
\vec{E}_{3}
\end{bmatrix}
\end{equation}
with
\begin{equation}
\mathbb{\mathbf{D}}_{\text{}}^{M}=
\left[\begin{array}{cc}
\bar{\pmb{\mathfrak{c}}}^M &
-\bar{\pmb{\mathfrak{e}}}^{M^T}\\
\bar{\pmb{\mathfrak{e}}}^M &
\bar{\tilde{\pmb{\mathfrak{e}}}}^M
\end{array}\right]
\end{equation}
where $\bar{\pmb{\mathfrak{c}}}^M$, $\bar{\pmb{\mathfrak{e}}}^M$, and $\bar{\tilde{\pmb{\mathfrak{e}}}}^M$
are the matrices of the secant elastic, piezoelectric and dielectric coefficients.
The internal energy function per unit volume for the solid is: \color{black}
\begin{equation}
\Pi_{int}^{M,sol}=\frac{1}{2}{\mathbf{E}_{g}}^{T}\text{ }\mathbf{D}^{M}(\mathbf{E}_{g})\text{ }{\mathbf{E}_{g}}
\end{equation} \color{black}
where the superscript $sol$ stands for solid\color{black}. If $V$ now indicates the total shell volume $V_{shell}$,
the potential of the internal forces acting on the shell can be obtained
through integration as:
\begin{eqnarray}
\Pi_{int}^{M,shell}=\int_{V}\Pi_{int}^{M,sol}\text{dV}=\frac{1}{2}\int_{V}{\mathbf{E}_{g}}^{T}\text{ }\mathbf{D}^{M}\text{ }{\mathbf{E}_{g}}\text{dV}=\frac{1}{2}\int_{V}{\mathbf{E}_{s}}^{T}\mathbb{\mathbb{\mathbf{A}}}^{T}\mathbf{D}^{M}\mathbb{\mathbb{\mathbb{\mathbf{A}}}\,}{\mathbf{E}_{s}}\text{dV}\nonumber \\
=\frac{1}{2}\int_{A}{\mathbf{E}_{s}}^{T}\left[h\int_{-\frac{1}{2}}^{\frac{1}{2}}\mathbb{\mathbb{\mathbf{A}}}^{T}\mathbf{D}^{M}\mathbb{\mathbb{\mathbb{\mathbf{A}}}}\bar {\mu}\text{d\ensuremath{\xi}}^{3}\right]{\mathbf{E}_{s}}\text{dA}\label{eq:pi_shell_0}\\
=\frac{1}{2}\int_{A}{\mathbf{E}_{s}}^{T}\mathbf{D}_{Macro}^{Shell}{\mathbf{E}_{s}}\text{dA}\nonumber
\end{eqnarray}
where eq. (\ref{eq36}) has been used and, in the last equality, the
macroscopic constitutive matrix of the shell $\mathbb{\textbf{D}}_{\text{Macro}}^{\text{Shell}}$
has been defined as \color{black}
\begin{equation}
\mathbb{\mathbf{D}}_{\text{Macro}}^{\text{Shell}}=h\int_{-\frac{1}{2}}^{\frac{1}{2}}\mathbb{\mathbf{A}}^{T}\mathbf{D}^{M}\mathbb{\mathbf{A}}\bar {\mu}\text{\ensuremath{\text{d\ensuremath{\xi}}^{3}}}\text{ }\label{eq:D_shell_macro}
\end{equation}
\color{black}
where $\bar {\mu}$ is the determinant of the shifter tensor that for slightly curved
shell is $\bar {\mu}\approx1$. \color{black}
On the other hand, the potential of the shell can be expressed as
\begin{equation}
\Pi_{int}^{M,shell}=\frac{1}{2}\int_{A}{\mathbf{E}_{s}}^{T}\bar{\mathbf{L}}\text{dA}\label{eq:pi_shell}
\end{equation}
where $\bar{\mathcal{\mathbf{L}}}$ is the vector of stress resultants
of the shell given by

\begin{equation}
\bar{\mathcal{\mathbf{L}}}=\left[n_{11},n_{22},n_{12},m_{11},m_{22},m_{12},p_{1},p_{2},-\vec{d}_{1},-\vec{d}_{2},n^{0}{}_{33},n^{1}{}_{33},-\vec{d}^{0}{}_{3},-\vec{d}^{1}{}_{3}\right]^{T}
\end{equation}
where $n_{\alpha\beta}$ are the membrane forces, $m_{\alpha\beta}$
the bending moments, $p_{\alpha}$ the shear forces, $\vec{d}_{\alpha}$
the electric displacements, whereas $n^{0}{}_{33},n^{1}{}_{33}$,
$-\vec{d}^{0}{}_{3},-\vec{d}^{1}{}_{3}$ are the constant and linear
parts of the components in the thickness direction. Therefore, the
comparison between eqs. (\ref{eq:pi_shell_0}) and (\ref{eq:pi_shell})
leads to the expression of $\bar{\mathcal{\mathbf{L}}}$ as

\begin{equation}
\bar{\mathcal{\mathbf{L}}}=\mathbb{\mathbf{D}}_{\text{Macro}}^{\text{Shell}}\mathbf{E}_{s}
\end{equation}

\section{{\normalsize{}Results}}
\color{black}
As follows, the presented computational procedure is applied to a
simple test case, whereby an RVE containing four PVDF fibers in a
square configuration is considered. The final aim is to predict the effective properties of the homogenized shell element. Each fiber is considered as a linear piezoelastic solid and is discretized with linear 8-node brick elements (Fig. \ref{fig:11}).
Similarly to \cite{key-18} the nonlinear theory is specialized to the kinematically linear case where the mechanical
constitutive behavior is defined by the following constitutive law:
\begin{equation}\label{888}
\pmb{\sigma}=\frac{E}{1+\nu}\left(\frac{\nu}{1-2\nu}tr(\pmb{\epsilon})\textbf{1}+\pmb{\epsilon}\right)
\end{equation}
where $\pmb{\sigma}$ denotes the Cauchy stress, $\pmb{\epsilon}$ the linearized strain tensor with trace $tr(\pmb{\epsilon})$. Moreover E is the Young's modulus and $\nu$ is the Poisson's ratio. Frictionless electromechanical contact constraints are enforced at the interface between the fibers using the described electromechanical contact formulation. The final piezoelectric constitutive matrix of the PVDF material, $\mathbf{D}_{PVDF}^{m}$, simplifies to
\color{black}
\begin{equation}
\mathbf{D}_{PVDF}^{m}=\left(\begin{array}{ccccccccc}
\lambda+2\mu & \lambda & \lambda & 0 & 0 & 0 & 0 & 0 & -e_{31}\\
\lambda & \lambda+2\mu & \lambda & 0 & 0 & 0 & 0 & 0 & -e_{32}\\
\lambda & \lambda & \lambda+2\mu & 0 & 0 & 0 & 0 & 0 & -e_{33}\\
0 & 0 & 0 & \mu & 0 & 0 & 0 & 0 & 0\\
0 & 0 & 0 & 0 & \mu & 0 & 0 & 0 & 0\\
0 & 0 & 0 & 0 & 0 & \mu & 0 & 0 & 0\\
0 & 0 & 0 & 0 & 0 & 0 & \tilde{e}_{11} & 0 & 0\\
0 & 0 & 0 & 0 & 0 & 0 & 0 & \tilde{e}_{22} & 0\\
e_{31} & e_{32} & e_{33} & 0 & 0 & 0 & 0 & 0 & \tilde{e}_{33}
\end{array}\right)\label{100}
\end{equation}
where $\lambda$ and $\mu$ are the Lamé constants \cite{key-102}.

In this work it is assumed that  $\lambda$ and $\mu$ are equal to $80.3 N/mm^2$ and $58.1 N/mm^2$ (corresponding to $E=150N/mm^2$ and $\nu=0.3$), while the piezoelectric strain coefficients $d_{31}, d_{32}, d_{33}$ are $20 \cdot 10^{-12}, 3 \cdot 10^{-12}, -35 \cdot 10^{-12}$ m/V, respectively, and the permittivity coefficients  $\tilde{e}_{11}, \tilde{e}_{22}, \tilde{e}_{33}$ are all equal to $12 \cdot \varepsilon_{0}$, where $\varepsilon_{0}=8.854 \cdot 10^{-12} F/m$ is the Faraday constant.

\color{black}
 The fiber diameter ranges from 0.1$\mu$m up to 2$\mu$m according to applications and used process parameters. Moreover packing fibers in the horizontal and vertical direction we obtained fiber arrays at the macroscale with a thickness ranging from 1 up to 20$\mu$m. According to available experimental results, each fiber in the RVE is here assumed to have a radius R = 1.0$\mu$m, therefore each edge of the RVE has a length L=2R = 2.0$\mu$m.
\color{black}
To evaluate the macroscopic effective coefficients of an equivalent homogeneous solid, appropriate boundary conditions are repeatedly
applied to the unit cell in such a way that only one component of
the strain/electric field vector is non-zero at each time. Then each
effective coefficient can be easily determined by multiplying the
corresponding row of the material matrix $\mathbf{D}^{M}$ by the
strain/electric field vector. Table \ref{tab:1} lists the boundary
conditions that lead to the computation of all effective coefficients,
and illustration of the referenced RVE surfaces is reported in Fig.
\ref{fig:10}. \color{black}
For a transversely isotropic piezoelectric solid, the stiffness matrix,
the piezoelectric matrix and the dielectric matrix simplify, so that
there remain in all 11 independent coefficients. Thus, in the case
of aligned PVDF fibers the constitutive matrix for the homogenized
solid simplifies to
\begin{equation}
\mathbf{D}_{PVDF}^{M}=\begin{bmatrix}\bar{C}_{11} & \bar{C}_{12} & \bar{C}_{13} & 0 & 0 & 0 & 0 & 0 & -\bar{e}_{13}\\
\bar{C}_{12} & \bar{C}_{11} & \bar{C}_{13} & 0 & 0 & 0 & 0 & 0 & -\bar{e}_{13}\\
\bar{C}_{13} & \bar{C}_{13} & \bar{C}_{33} & 0 & 0 & 0 & 0 & 0 & -\bar{e}_{33}\\
0 & 0 & 0 & \bar{C}_{44} & 0 & 0 & 0 & -\bar{e}_{15} & 0\\
0 & 0 & 0 & 0 & \bar{C}_{44} & 0 & -\bar{e}_{15} & 0 & 0\\
0 & 0 & 0 & 0 & 0 & \bar{C}_{66} & 0 & 0 & 0\\
0 & 0 & 0 & 0 & \bar{e}_{15} & 0 & \bar{\tilde{e}}_{11} & 0 & 0\\
0 & 0 & 0 & \bar{e}_{15} & 0 & 0 & 0 & \bar{\tilde{e}}_{11} & 0\\
\bar{e}_{13} & \bar{e}_{13} & \bar{e}_{33} & 0 & 0 & 0 & 0 & 0 & \bar{\tilde{e}}_{33}
\end{bmatrix}\label{4.3}
\end{equation}
\color{black}
In particular for an hexagonal array packing the resulting composite is transversely
isotropic and for a square array it is tetragonal. \color{black}
Figs. \ref{fig:12} to \ref{fig:14} illustrate the contours of displacements,
stresses and electrical potential for some special cases of boundary
conditions applied to the RVE.

Based on $\mathbf{D}_{PVDF}^{M}$ in eq. \eqref{4.3}, after multiplication
and thickness integration as per eq. (\ref{eq:D_shell_macro}), the
expression for $\textbf{D}_{\text{Macro}}^{\text{Shell}}$ is obtained
as

\begin{equation} \resizebox{.9\hsize}{!}{$
\mathbf{D}_{Macro}^{Shell}=\left(\begin{array}{cccccccccccccc}h\bar{C}_{11} & h\bar{C}_{12} & 0 & 0 & 0 & 0 & 0 & 0 & 0 & 0 & h\bar{C}_{13} & 0 & -h\bar{e}_{13} & 0\\h\bar{C}_{12} & h\bar{C}_{11} & 0 & 0 & 0 & 0 & 0 & 0 & 0 & 0 & h\bar{C}_{13} & 0 & -h\bar{e}_{13} & 0\\0 & 0 & h\bar{C}_{44} & 0 & 0 & 0 & 0 & 0 & 0 & -h\bar{e}_{15} & 0 & 0 & 0 & 0\\0 & 0 & 0 & \frac{1}{12}h^{3}\bar{C}_{11} & \frac{1}{12}h^{3}\bar{C}_{12} & 0 & 0 & 0 & 0 & 0 & 0 & \frac{1}{12}h^{3}\bar{C}_{13} & 0 & -\frac{1}{12}h^{3}\bar{e}_{13}\\0 & 0 & 0 & \frac{1}{12}h^{3}\bar{C}_{12} & \frac{1}{12}h^{3}\bar{C}_{11} & 0 & 0 & 0 & 0 & 0 & 0 & \frac{1}{12}h^{3}\bar{C}_{13} & 0 & -\frac{1}{12}h^{3}\bar{e}_{13}\\0 & 0 & 0 & 0 & 0 & \frac{1}{12}h^{3}\bar{C}_{44} & 0 & 0 & 0 & 0 & 0 & 0 & 0 & 0\\0 & 0 & 0 & 0 & 0 & 0 & h\bar{C}_{44} & 0 & -h\bar{e}_{15} & 0 & 0 & 0 & 0 & 0\\0 & 0 & 0 & 0 & 0 & 0 & 0 & h\bar{C}_{66} & 0 & 0 & 0 & 0 & 0 & 0\\0 & 0 & 0 & 0 & 0 & 0 & h\bar{e}_{15} & 0 & h\bar{\tilde{e}}_{11} & 0 & 0 & 0 & 0 & 0\\0 & 0 & h\bar{e}_{15} & 0 & 0 & 0 & 0 & 0 & 0 & h\bar{\tilde{e}}_{11} & 0 & 0 & 0 & 0\\h\bar{C}_{13} & h\bar{C}_{13} & 0 & 0 & 0 & 0 & 0 & 0 & 0 & 0 & h\bar{C}_{33} & 0 & -h\bar{e}_{33} & 0\\0 & 0 & 0 & \frac{1}{12}h^{3}\bar{C}_{13} & \frac{1}{12}h^{3}\bar{C}_{13} & 0 & 0 & 0 & 0 & 0 & 0 & \frac{1}{12}h^{3}\bar{C}_{33} & 0 & -\frac{1}{12}h^{3}\bar{e}_{33}\\h\bar{e}_{13} & h\bar{e}_{13} & 0 & 0 & 0 & 0 & 0 & 0 & 0 & 0 & h\bar{e}_{33} & 0 & h\bar{\tilde{e}}_{33} & 0\\0 & 0 & 0 & \frac{1}{12}h^{3}\bar{e}_{13} & \frac{1}{12}h^{3}\bar{e}_{13} & 0 & 0 & 0 & 0 & 0 & 0 & \frac{1}{12}h^{3}\bar{e}_{33} & 0 & \frac{1}{12}h^{3}\bar{\tilde{e}}_{33}\end{array}\right)
$} \end{equation}

\color{black} Eq. 81 is derived here assuming $\mathbf{D}_{PVDF}^{M}$ constant during thickness integration.
Since $\mathbf{D}_{PVDF}^{M}$ represents an averaged secant matrix obtained taking into account the local variation of the deformation
$\mathbf{E}_{g}$ in an RVE of thickness h, this is a valid approximation if the PVDF fibers arrangement does not change in
the thickness, otherwise further Gauss points with different RVE geometry are required leading to different $\mathbf{D}_{PVDF}^{M}$
for each layer. \color{black}
Following this procedure, the evolution of selected coefficients $\left(D_{\text{Macro}}^{\text{Shell}}\right){}_{ij}$
with the increase of the prescribed boundary conditions is reported
in Fig. \ref{fig:15}.
The observed nonlinear behavior is due to electromechanical contact
between the fibers at the microscale.

The proposed computational technique can be easily extended to determine
homogenized coefficients for more complex RVE geometries, such as
for RVEs with fibers of arbitrary cross-section or different fiber
arrangements.

\section{{\normalsize{}Conclusions}}
This paper focused on the development of a numerical strategy to predict
the behavior of flexible piezoelectric materials and devices made
of polymeric nanofibers. These materials are attractive for several
technological applications including nanogenerators, mechanical energy
harvesting and pressure/force sensors. The objective was to establish a
link between the macroscopic performance of the sheets and their microscale
features, i.e. the detailed fiber geometry and material properties,
additionally accounting for the electromechanical contact interactions
between the fibers. A simple multiscale and multiphysics computational procedure was
set up for this purpose. The kinematically nonlinear electro-elasticity theory was introduced
starting from a variational setting. Consistent linearization was performed with the automatic differentiation technique. A two-step homogenization procedure was then adopted to derive the macroscopic constitutive behavior of the shell based on
the results of the microscopic boundary value problem. An example
was presented to demonstrate the feasibility of the proposed strategy.
In particular the developed procedure highlights the nonlinear
behaviour of the homogenized material and the different response
under tension and compression loads. Future extensions will include the set up of a fully coupled computational procedure in the spirit of the FE$^{2}$ method, as well as the comparison with available experimental results.
\color{black}The availability of such concurrent multiscale-multiphysics framework, upon experimental validation,
will ultimately enable the optimal microstructural design of thin flexible piezoelectric devices and
thus the tailoring of their macroscopic performance.
\color{black}
\section{{\normalsize{}Acknowledgements}}

Claudio Maruccio acknowledges the support from the Italian MIUR through
the project FIRB Futuro in Ricerca 2010 \textit{Structural mechanics
models for renewable energy applications} (RBFR107AKG). Laura De
Lorenzis, Dario Pisignano and Luana Persano acknowledge the support
from the European Research Council under the European Union's Seventh
Framework Programme (FP7/2007-2013), ERC Starting Grants INTERFACES
(L. De Lorenzis, grant agreement n. 279439) and NANO-JETS (D. Pisignano
and L. Persano, grant agreement n. 306357). \color{black}Furthermore, the authors gratefully
acknowledge the reviewers for useful comments and suggestions that contributed to improve
the quality of the paper. \color{black}

\begin{center}
\begin{table}[t]
\begin{centering}
\begin{tabular}{c|c|c|c|c|c|c|c}
\hline
Eff. Coeff. & $A^{-}$ & $A^{+}$ & $B^{-}$ & $B^{+}$ & $C^{-}$ & $C^{+}$ & Formula\tabularnewline
\hline
\hline
 & $u_{i}/\phi$ & $u_{i}/\phi$ & $u_{i}/\phi$ & $u_{i}/\phi$ & $u_{i}/\phi$ & $u_{i}/\phi$ & \tabularnewline
\hline
$\bar{C}_{11}$ & 0/- & $u_{1}$/- & 0/- & 0/- & 0/0 & 0/0 & $\bar{\sigma}_{11}/\bar{\epsilon}_{11}$\tabularnewline
\hline
$\bar{C}_{12}$ & 0/- & $u_{1}$/- & 0/- & 0/- & 0/0 & 0/0 & $\bar{\sigma}_{22}/\bar{\epsilon}_{11}$\tabularnewline
\hline
$\bar{C}_{13}$ & 0/- & 0/- & 0/- & 0/- & 0/0 & $u_{3}$/- & $\bar{\sigma}_{11}/\bar{\epsilon}_{33}$\tabularnewline
\hline
$\bar{C}_{33}$ & 0/- & 0/- & 0/- & 0/- & 0/0 & $u_{3}$/- & $\bar{\sigma}_{33}/\bar{\epsilon}_{33}$\tabularnewline
\hline
$\bar{C}_{44}$ & ($u_{3}$)/0 & ($u_{3}$)/0 & 0/- & 0/- & ($u_{1}$)/- & ($u_{1}$)/- & $\bar{\sigma}_{13}/\bar{\epsilon}_{31}$\tabularnewline
\hline
$\bar{C}_{66}$ & ($u_{2}$)/- & ($u_{2}$)/- & ($u_{1}$)/- & ($u_{1}$)/- & 0/0 & 0/0 & $\bar{\sigma}_{12}/\bar{\epsilon}_{12}$\tabularnewline
\hline
$\bar{e}_{13}$ & 0/- & 0/- & 0/- & 0/- & 0/0 & 0/$\phi$ & $-\bar{\sigma}_{11}/\bar{\vec{E}}_{3}$\tabularnewline
\hline
$\bar{e}_{33}$ & 0/- & 0/- & 0/- & 0/- & 0/0 & 0/$\phi$ & $-\bar{\sigma}_{33}/\bar{\vec{E}}_{3}$\tabularnewline
\hline
$\bar{e}_{15}$ & ($u_{3}$)/0 & ($u_{3}$)/0 & 0/- & 0/- & ($u_{1}$)/- & ($u_{1}$)/- & $\bar{\vec{D}}_{1}/\bar{\epsilon}_{31}$\tabularnewline
\hline
$\bar{\tilde{e}}_{11}$$ $ & 0/0 & 0/$\phi$ & 0/- & 0/- & 0/- & 0/- & $\bar{\vec{D}}_{1}/\bar{\vec{E}}_{1}$\tabularnewline
\hline
$\bar{\tilde{e}}_{33}$$ $ & 0/- & 0/- & 0/- & 0/- & 0/0 & 0/$\phi$ & $\bar{\vec{D}}_{3}/\bar{\vec{E}}_{3}$\tabularnewline
\hline
\end{tabular}
\par\end{centering}

\caption{{\small{}\label{tab:1}Boundary conditions and equations for the calculation
of the effective coefficients (see also Fig. \ref{fig:10}).}}
\end{table}

\par\end{center}

\noindent \begin{center}
\begin{figure}[!h]
\centering{}\includegraphics[height=75mm]{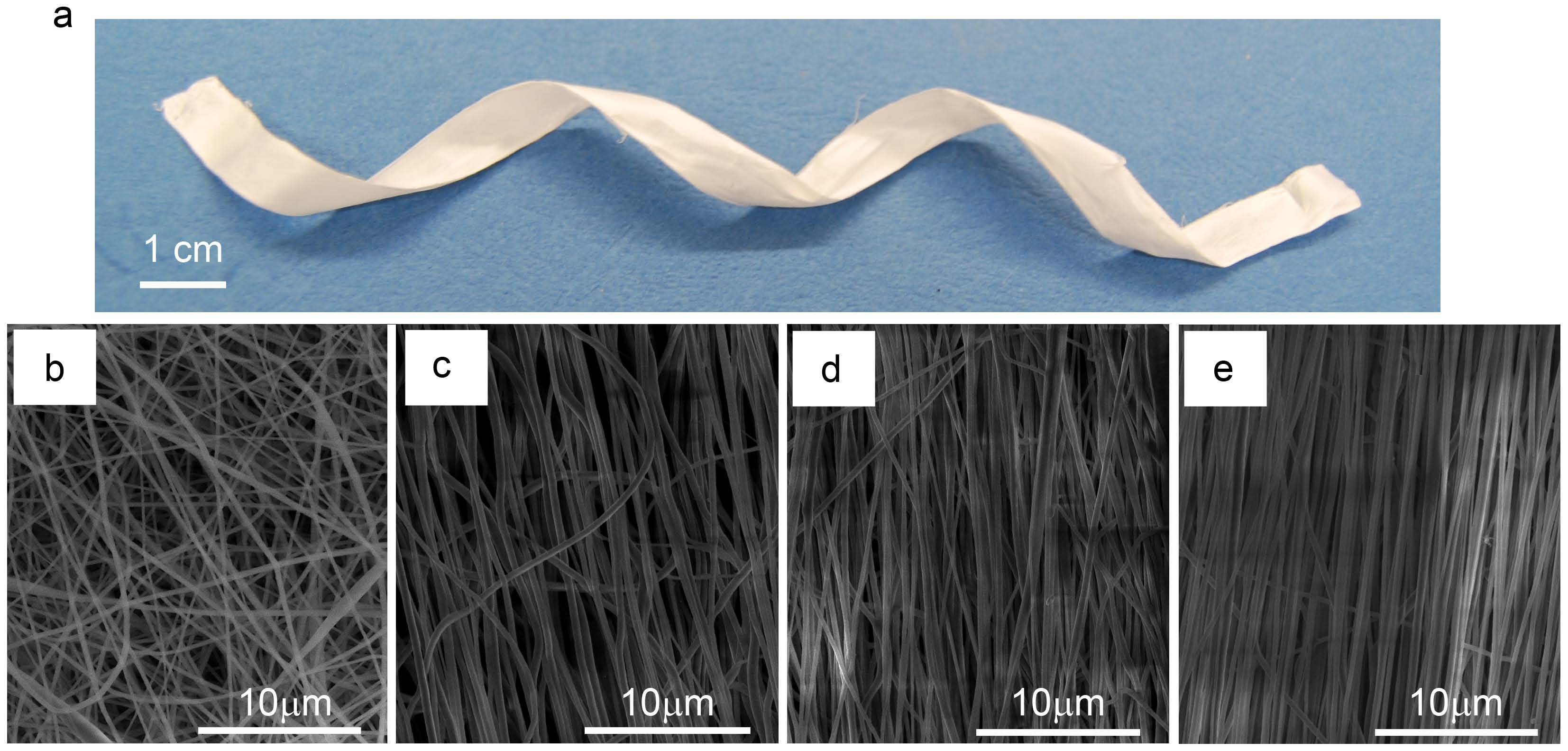} \caption{{\small{}\label{fig:1}(a) Photograph of a free-standing sheet of
aligned piezoelectric nanofibers made by electrospinning. (b-e) Scanning
electron micrographs of fiber architectures at microscale, ranging
from a fully random geometry (a) to excellent mutual alignment (e)
depending on the use process parameter. These nanofibers are realizing
by using a PVDF-based solution with a polymer concentration of about
1/5 (w/w) in dimethylformamide/acetone (3/2 v/v), and by using either
static metal collecting surface (a) or a disk collector rotating at
increasing angular speeds up to 4000 rpm (b-e). }}
\end{figure}

\par\end{center}

\begin{center}
\begin{figure}[!h]
\centering{}\includegraphics[height=95mm]{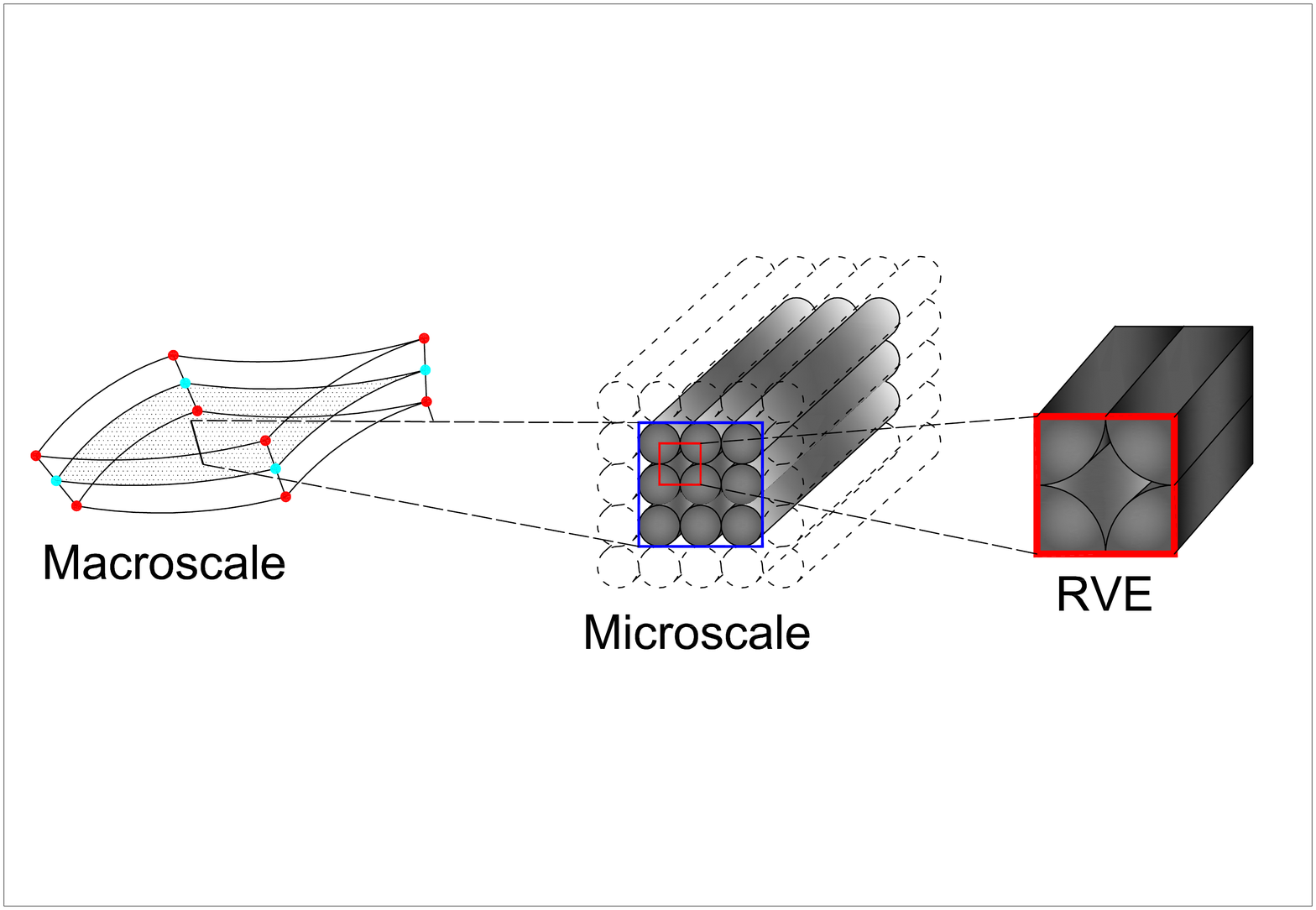} \caption{{\small{}\label{fig:6}Developed computational homogenization scheme
for piezoelectric thin sheets. Macroscale, microscale and RVE representation.}}
\end{figure}

\par\end{center}

\begin{center}
\begin{figure}[!h]
\centering{}\includegraphics[height=95mm]{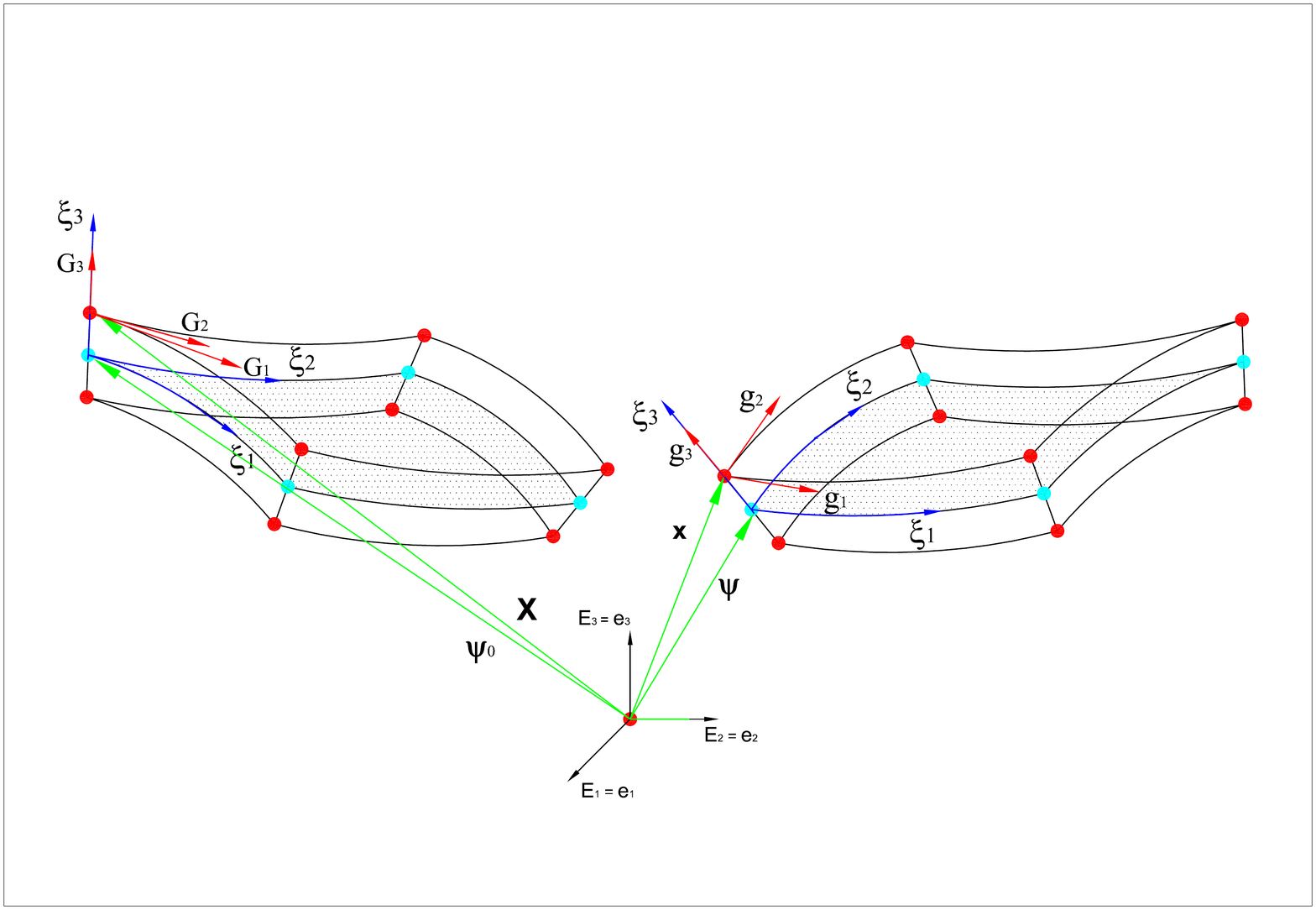} \caption{{\small{}\label{fig:7}Shell kinematics.}}
\end{figure}

\par\end{center}

\noindent \begin{center}
\begin{figure}[!h]
\centering{}\includegraphics[height=95mm]{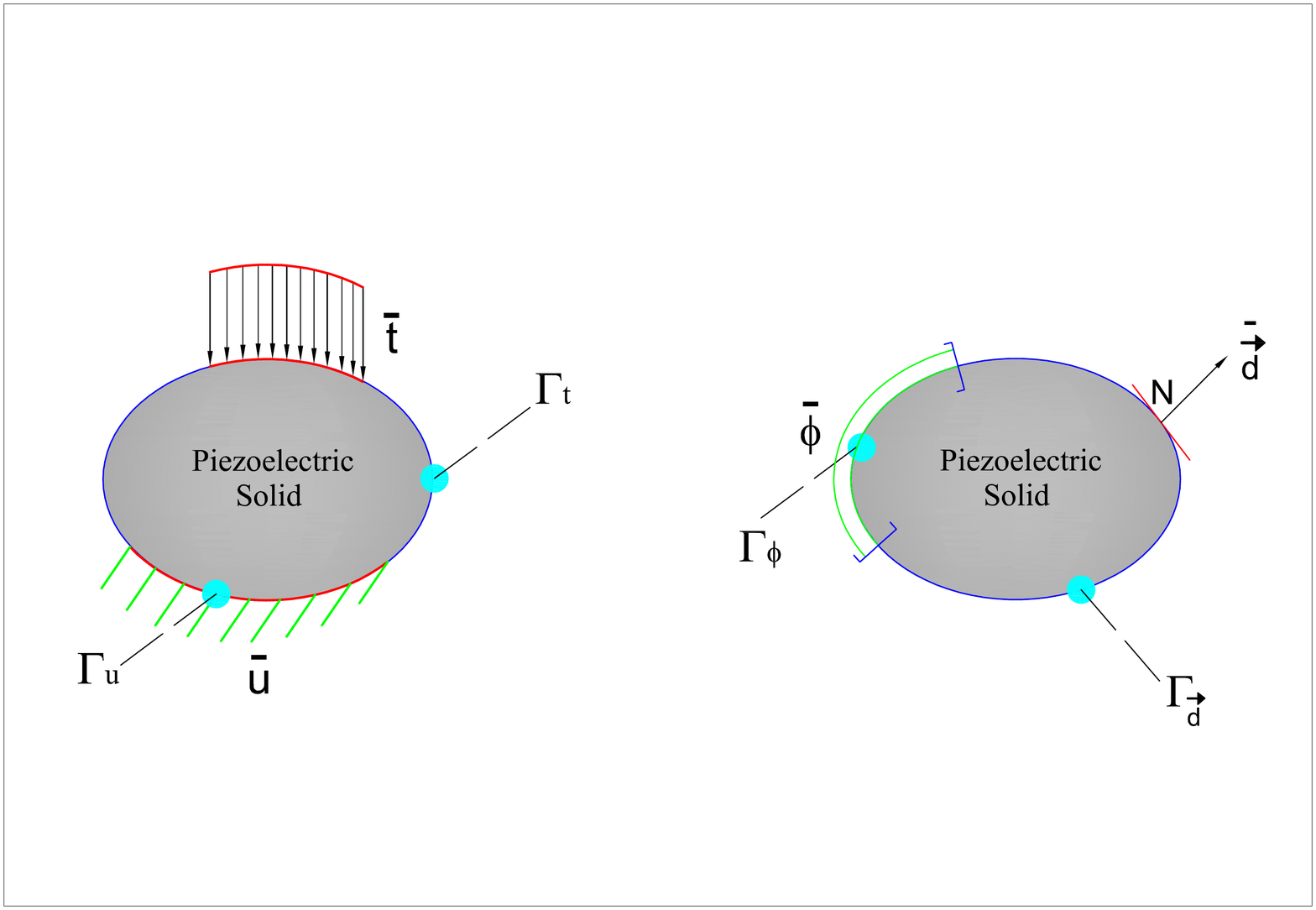} \caption{{\small{}\label{fig:7b}Domain boundary decomposition in mechanical
and electrical parts.}}
\end{figure}

\par\end{center}

\noindent \begin{center}
\begin{figure}[!h]
\centering{}\includegraphics[angle=90,height=95mm]{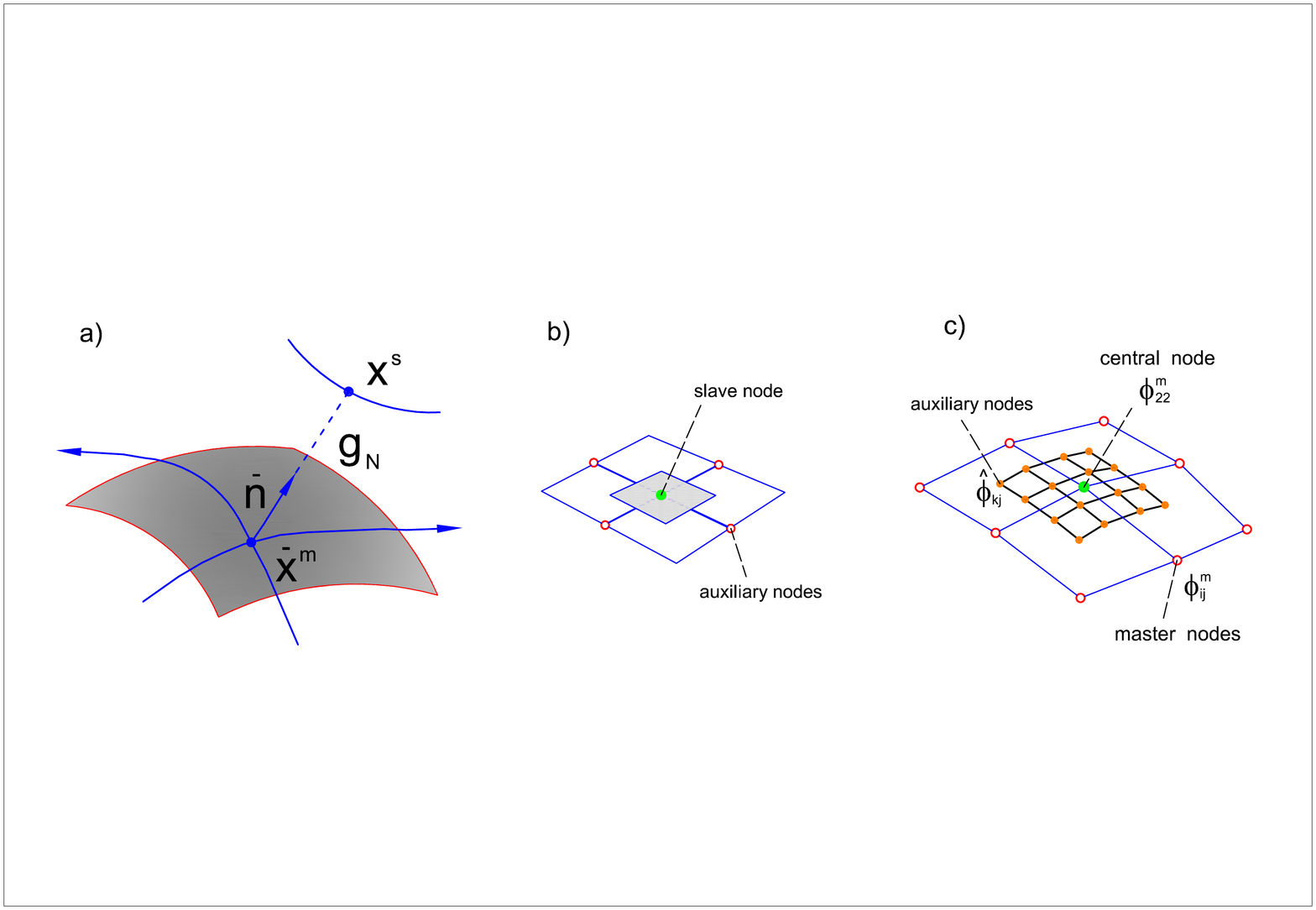} \caption{{\small{}\label{fig:7c}a) Closest point projection of the slave point
on the master surface b) slave side discretization c) smooth master
side discretization.}}
\end{figure}

\par\end{center}

\noindent \begin{center}
\begin{figure}[!h]
\centering{}\includegraphics[height=95mm]{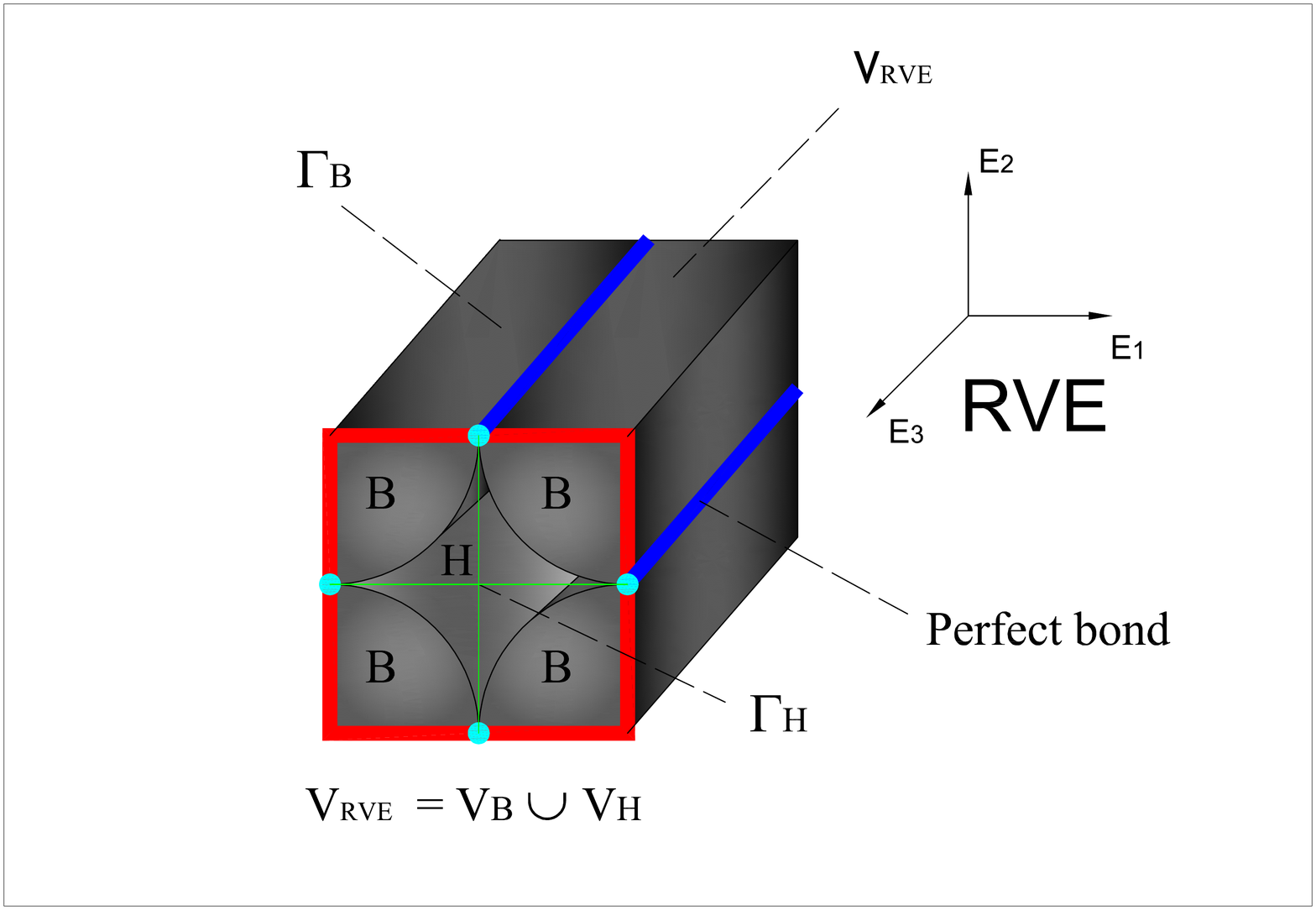} \caption{{\small{}\label{fig:8}RVE geometry.}}
\end{figure}

\par\end{center}

\begin{center}
\begin{figure}[!h]
\centering{}\includegraphics[height=95mm]{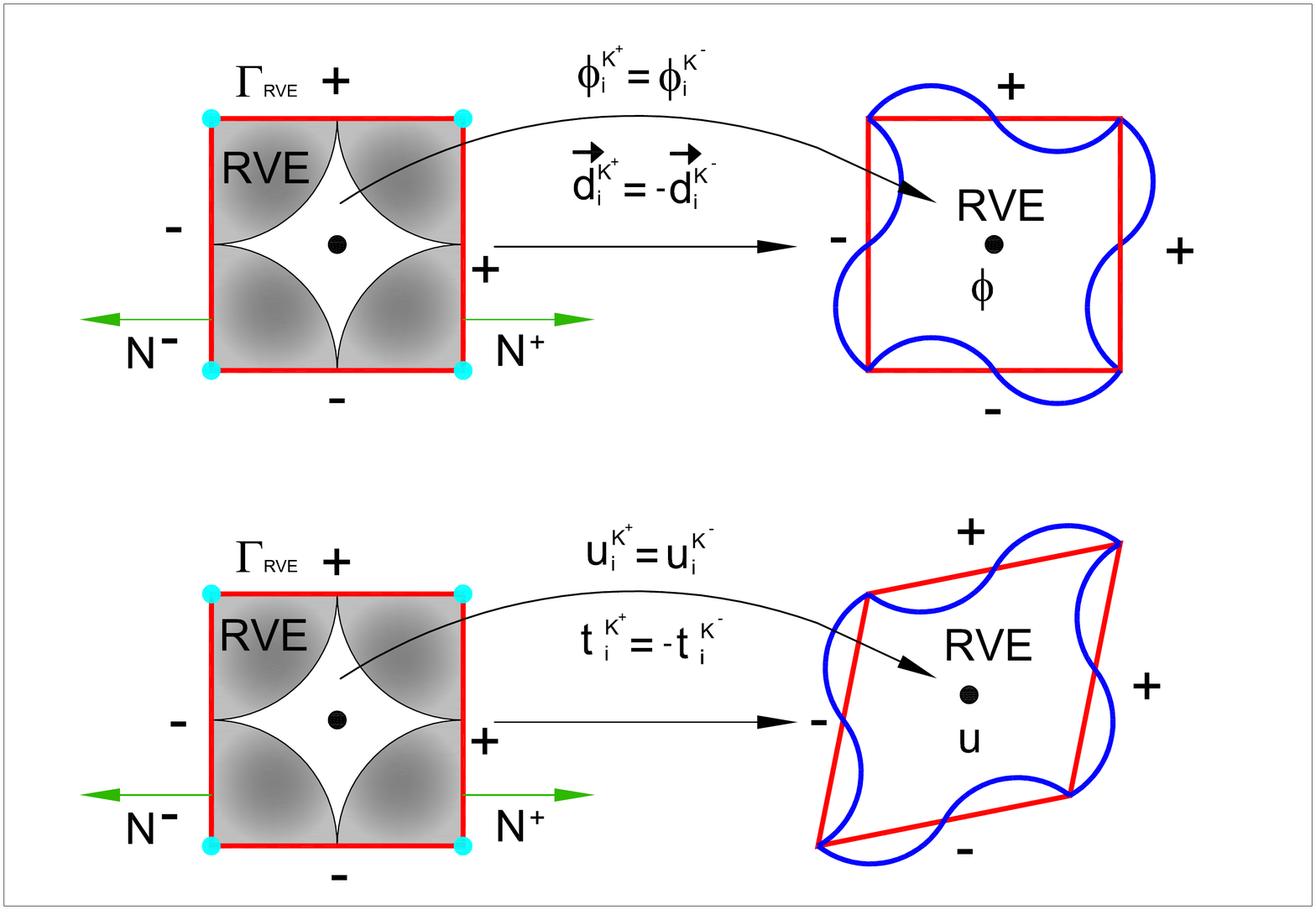} \caption{{\small{}\label{fig:9}Periodic boundary conditions.}}
\end{figure}

\par\end{center}

\begin{center}
\begin{figure}[!h]
\centering{}\includegraphics[height=95mm]{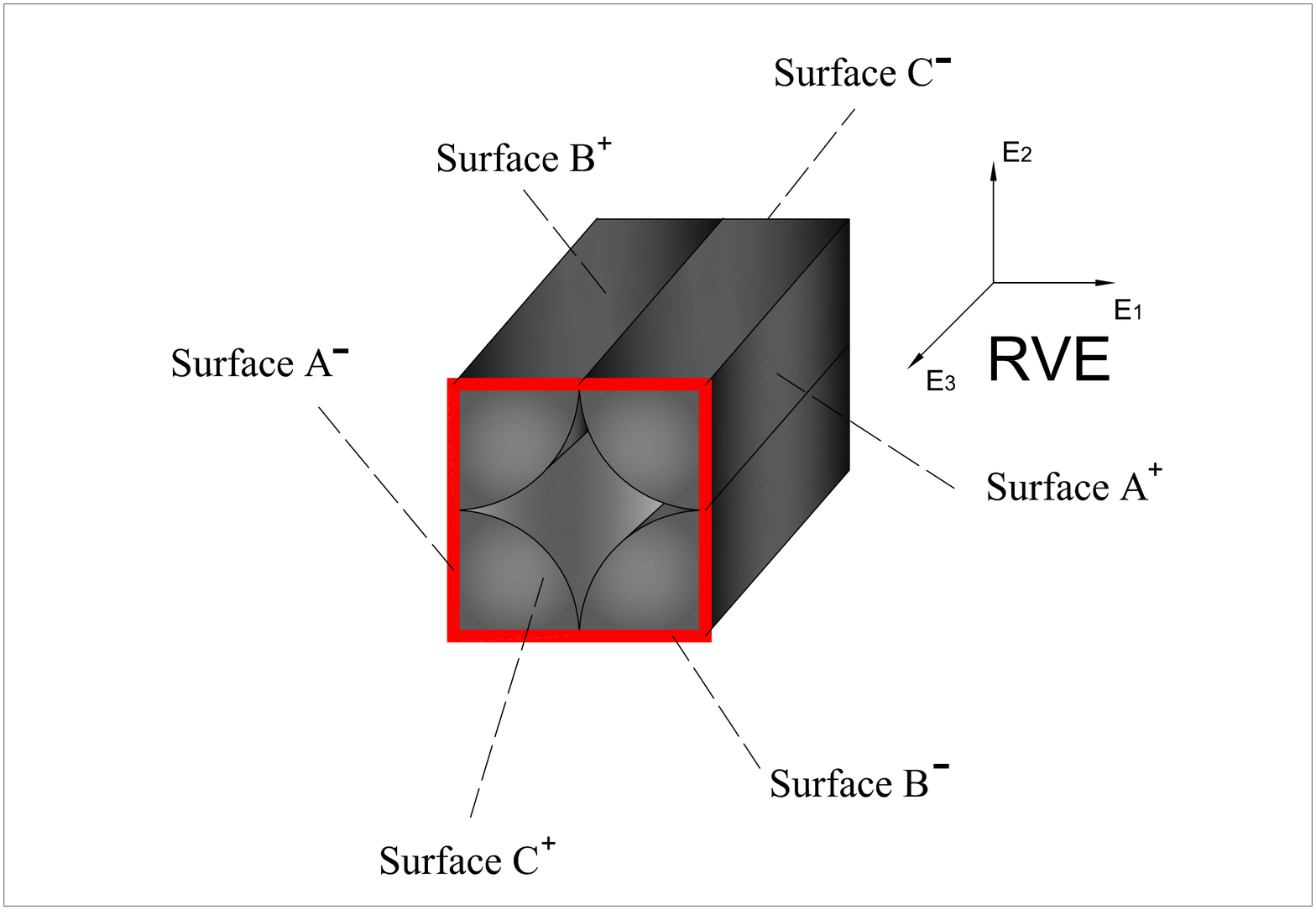} \caption{{\small{}\label{fig:10}RVE, definition of simple loading conditions.}}
\end{figure}

\par\end{center}

\begin{center}
\begin{figure}[!h]
\centering{}\includegraphics[height=55mm]{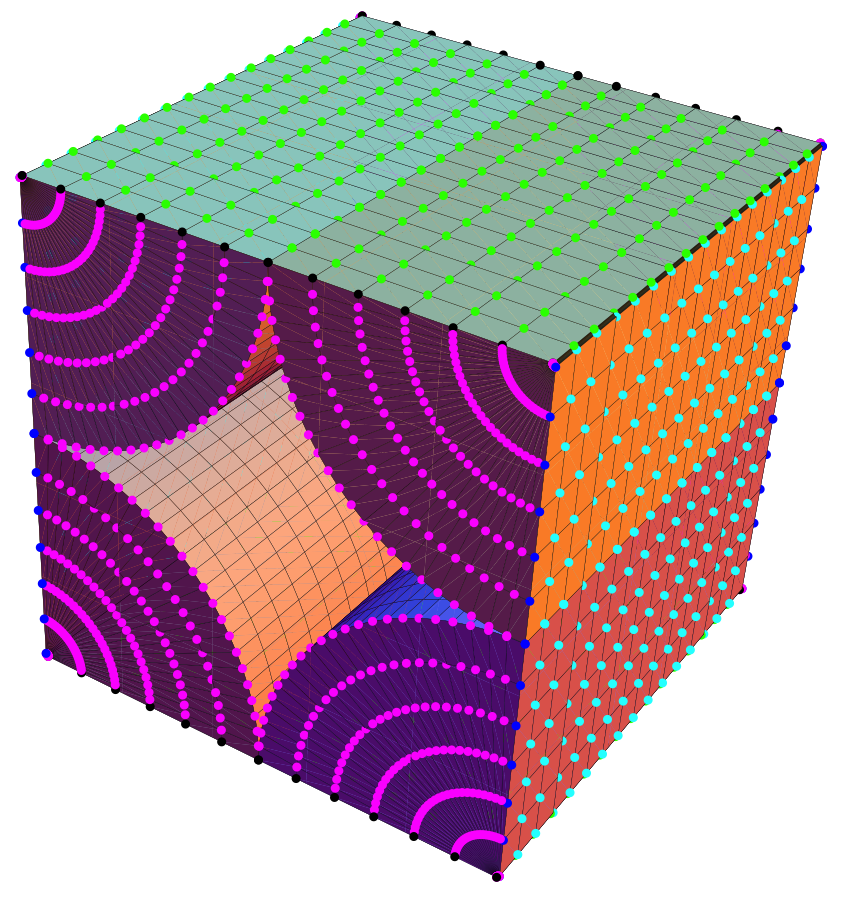} \caption{{\small{}\label{fig:11}RVE mesh and boundary conditions.}}
\end{figure}

\par\end{center}

\begin{center}
\begin{figure}[!h]
\noindent \begin{centering}
\subfloat[$u{}_{1}$]{\begin{centering}
\includegraphics[height=50mm]{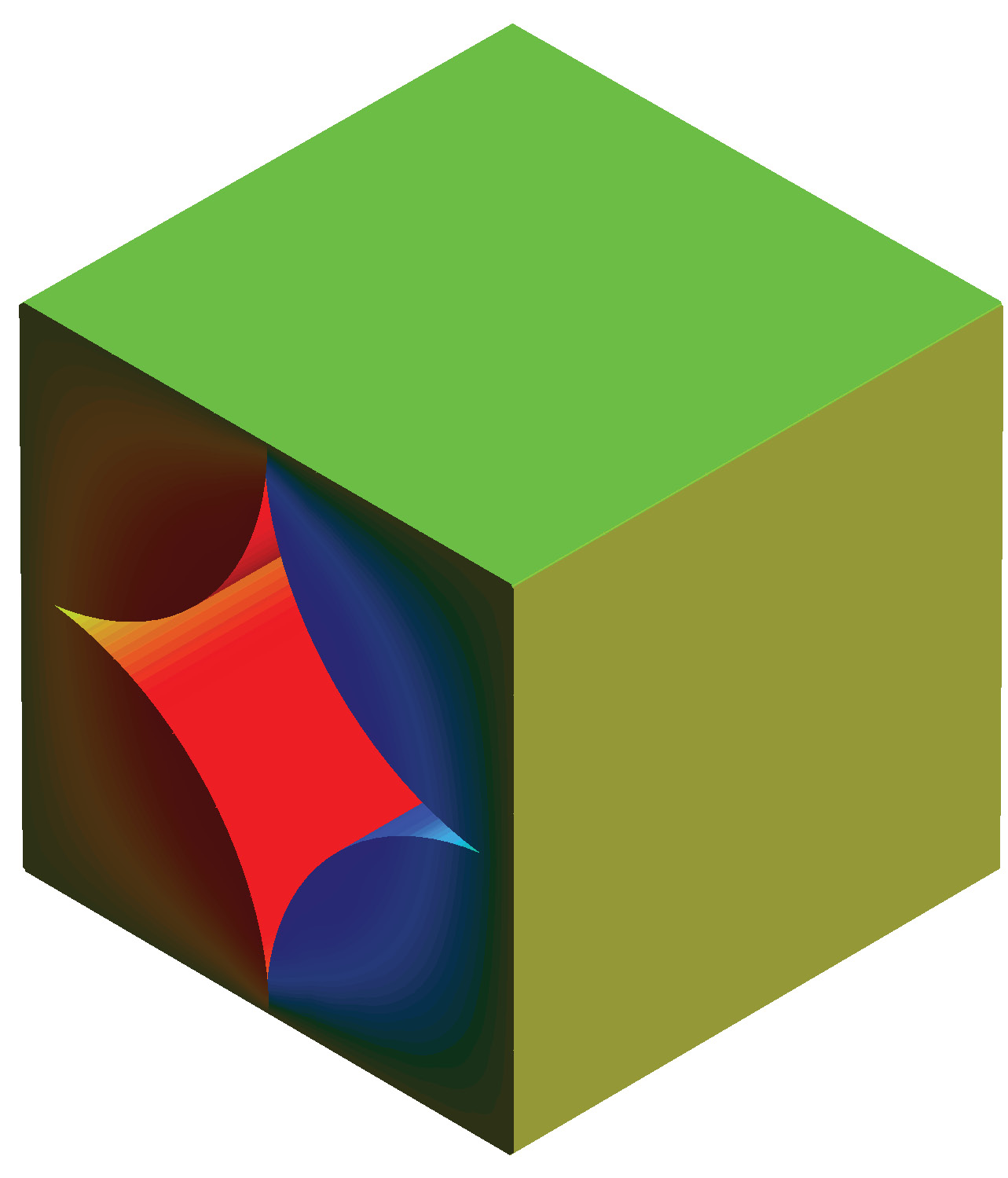}
\par\end{centering}

}\subfloat[$u{}_{2}$]{\begin{centering}
\includegraphics[height=50mm]{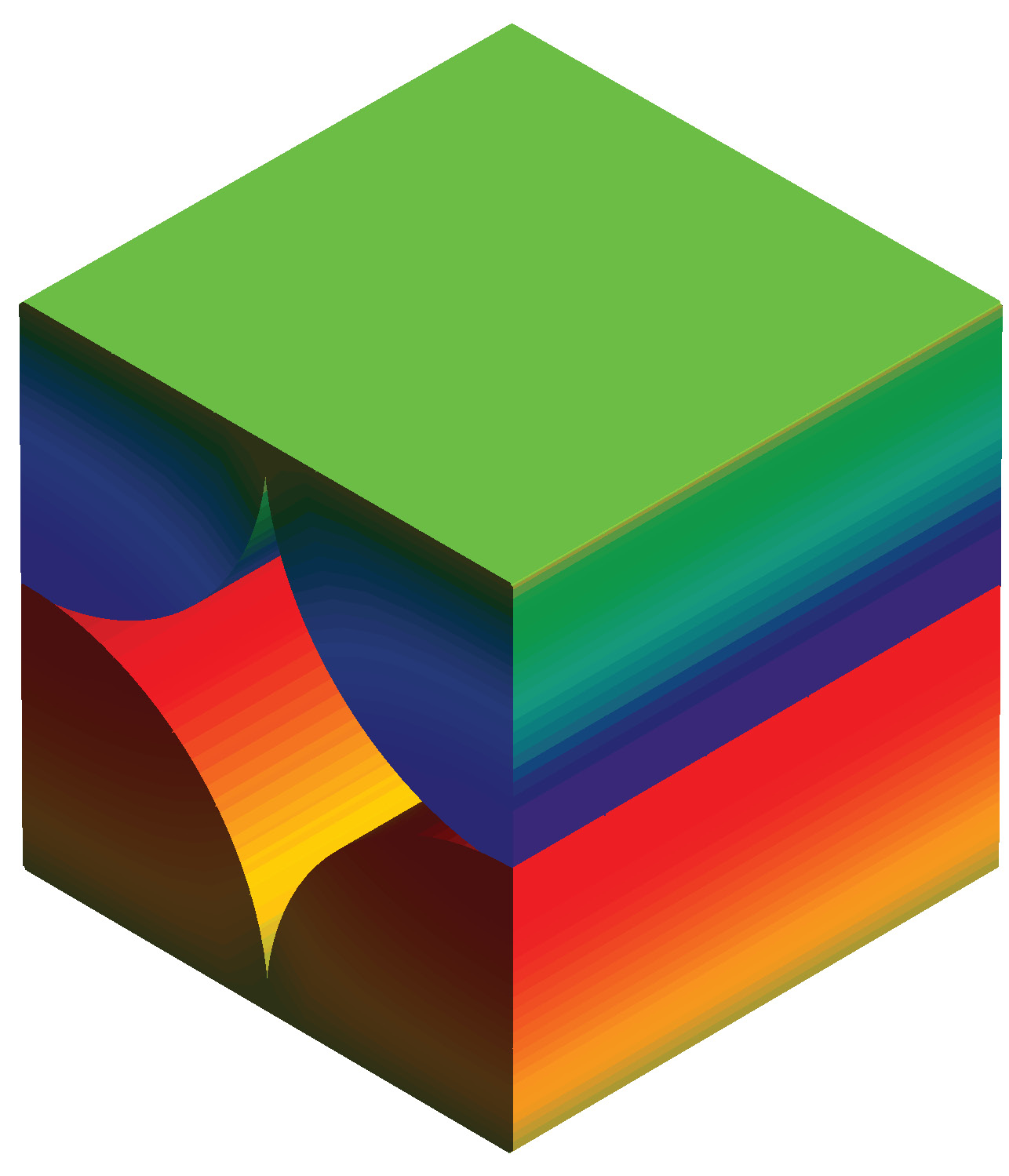}
\par\end{centering}

}
\par\end{centering}

\noindent \begin{centering}
\subfloat[$u{}_{3}$]{\begin{centering}
\includegraphics[height=50mm]{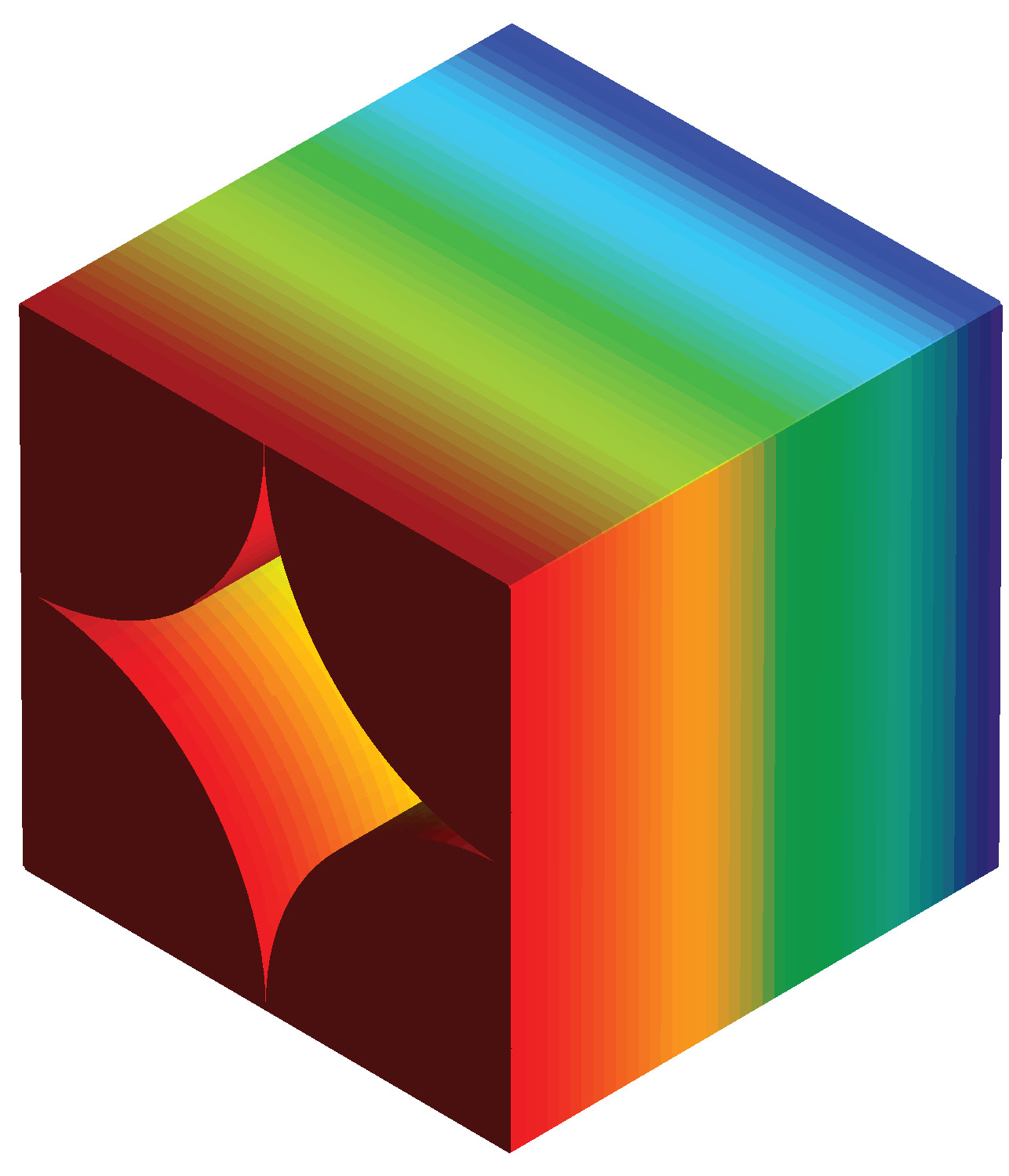}
\par\end{centering}

}\subfloat[$axes$]{\begin{centering}
\includegraphics[height=40mm]{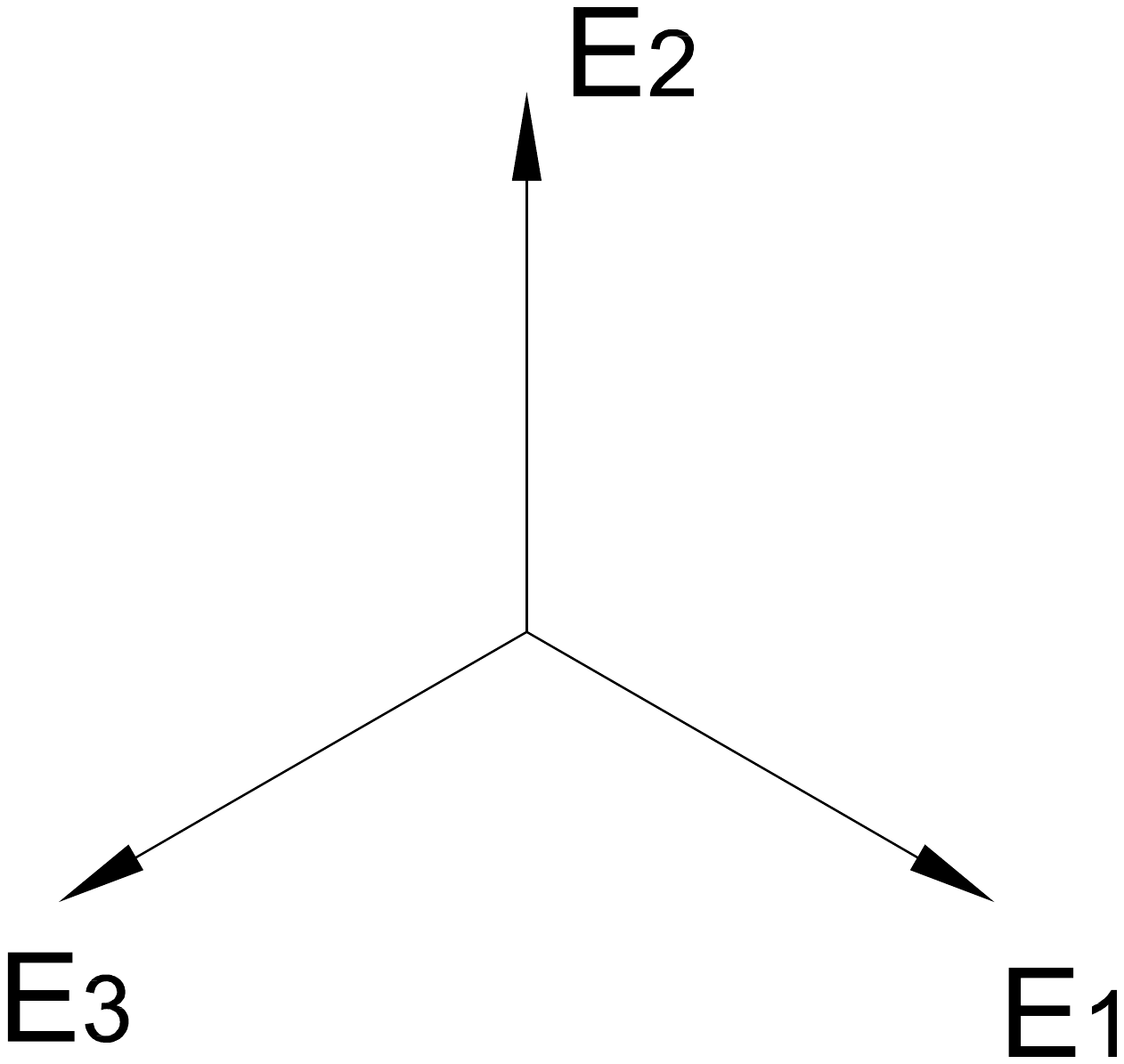}
\par\end{centering}
\color{white}
}\subfloat[]{\begin{centering}
\includegraphics[height=40mm]{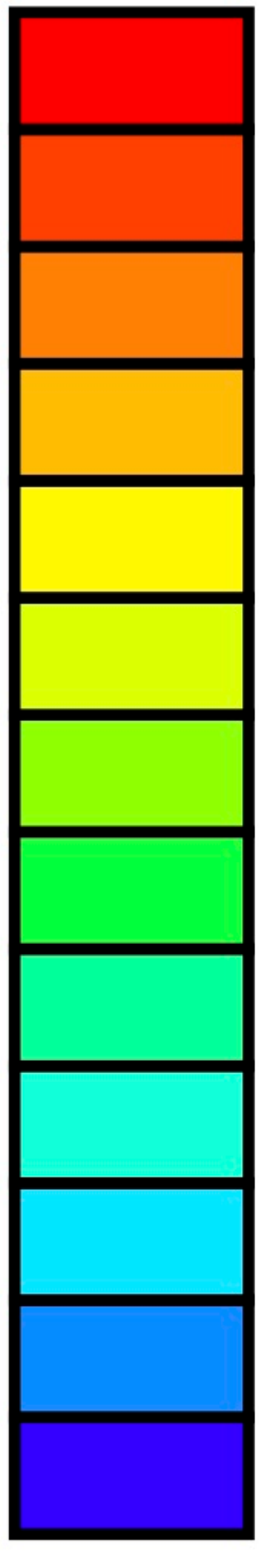}
\par\end{centering}

}
\par\end{centering}
\color{black}
\centering{}\caption{{\small{}\label{fig:12}Microscale behavior, countour levels of displacement
distribution in the fibers for a compression load along $E_3$. Maximum and minimum corresponding to the vertical color scales are: a) +/- 0.104 {[}}$\mu$m{\small{}{]}; b) +/- 0.18 {[}}$\mu$m{\small{}{]}; c) -0.1/0 {[}}$\mu$m{\small{}{]}.}}
\end{figure}

\par\end{center}

\begin{center}
\begin{figure}[!h]
\noindent \begin{centering}
\subfloat[$\sigma{}_{11}$]{\begin{centering}
\includegraphics[height=50mm]{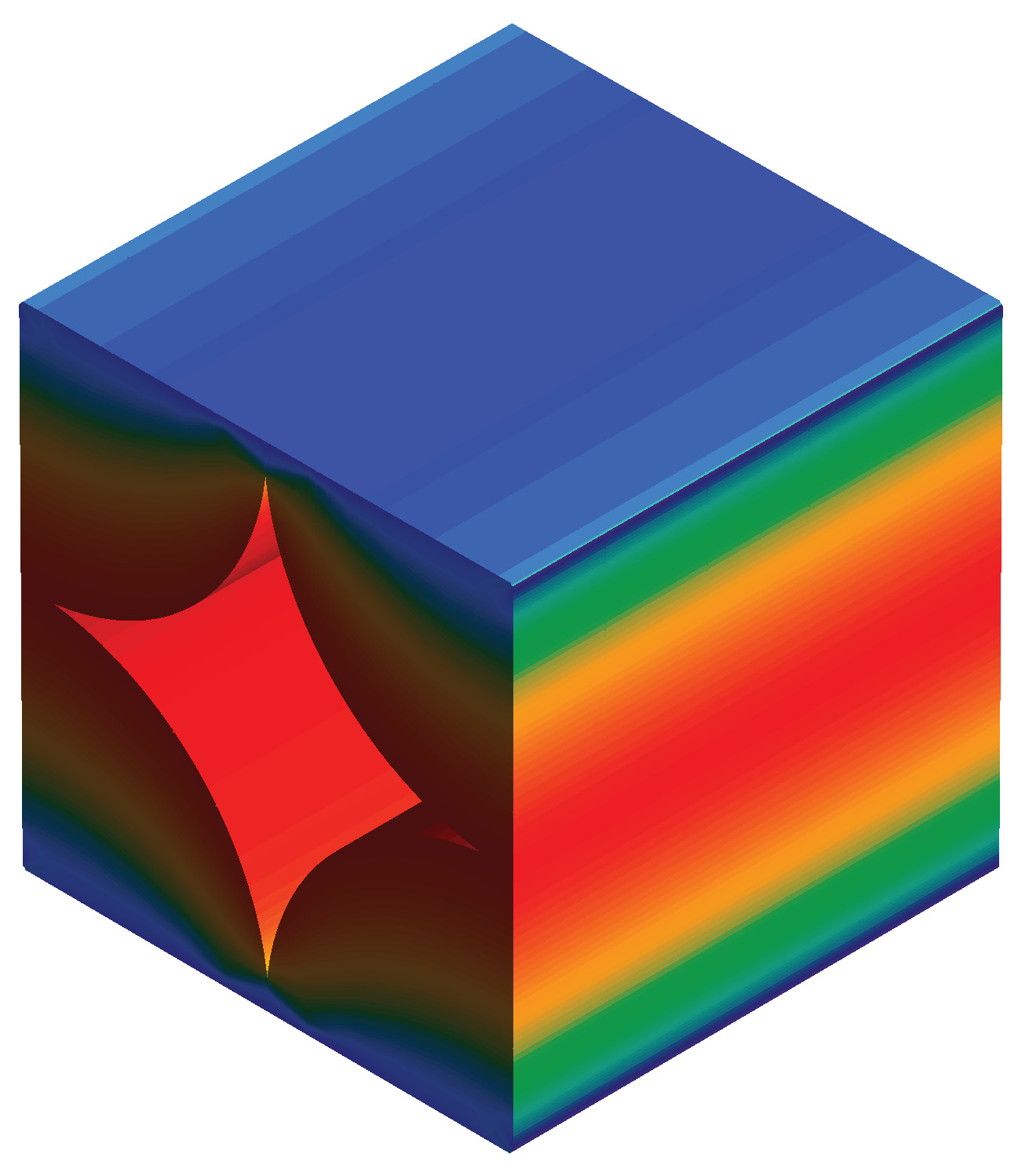}
\par\end{centering}

}\subfloat[$\sigma{}_{22}$]{\begin{centering}
\includegraphics[height=50mm]{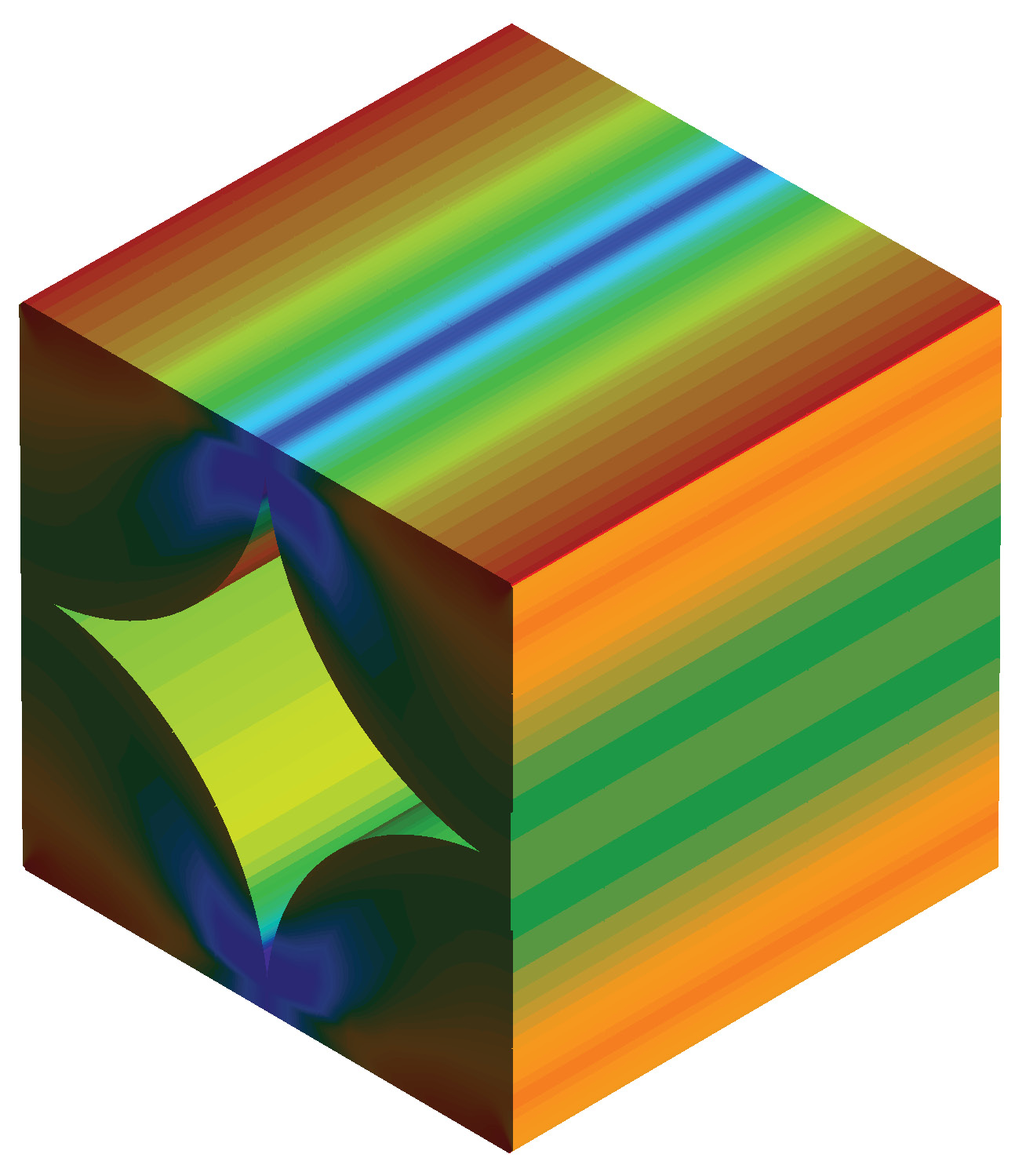}
\par\end{centering}

}\subfloat[$\sigma{}_{33}$]{\begin{centering}
\includegraphics[height=50mm]{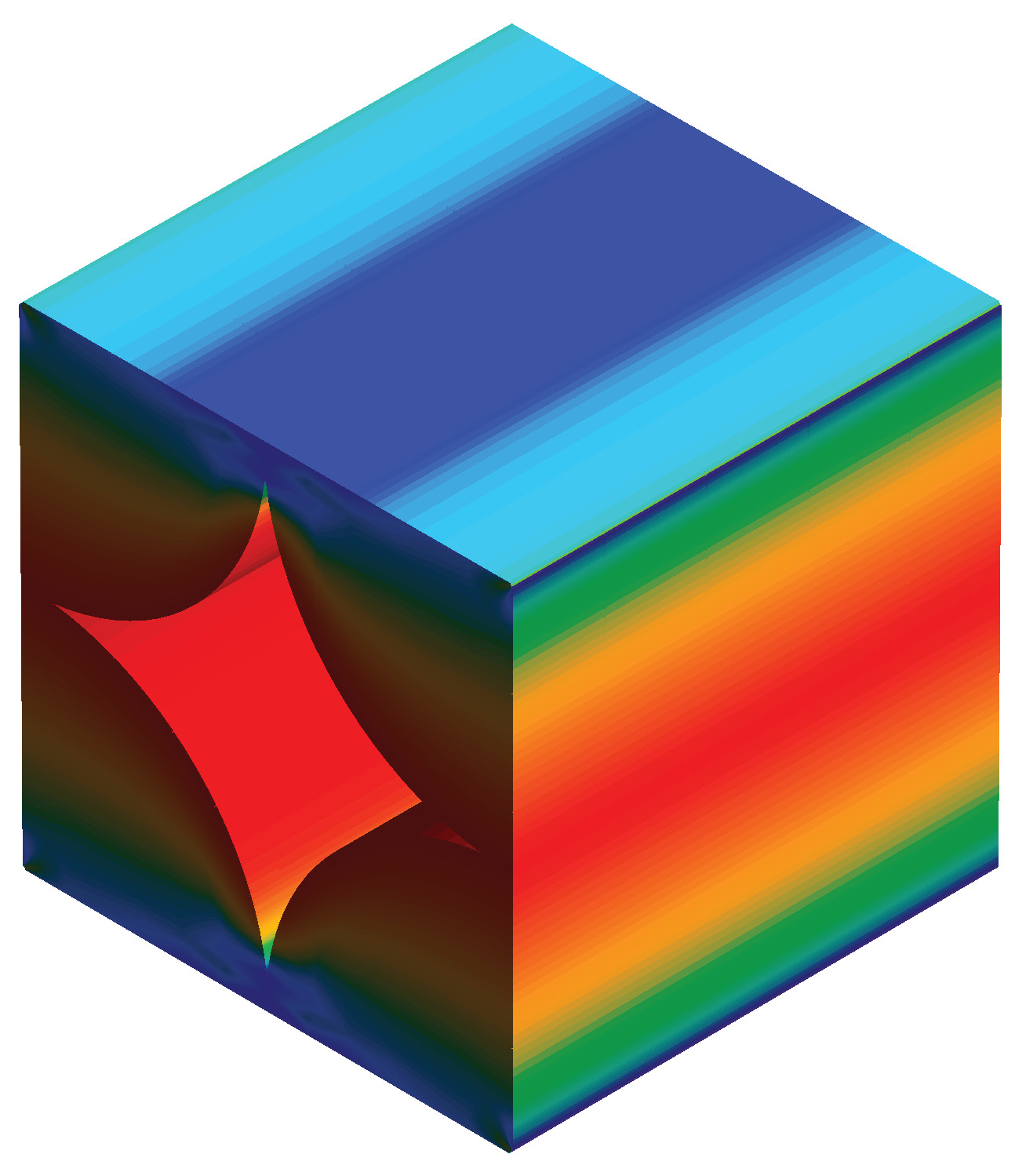}
\par\end{centering}

}\subfloat[$axes$]{\begin{centering}
\includegraphics[height=40mm]{SR1}
\par\end{centering}

\color{white}
}\subfloat[]{\begin{centering}
\includegraphics[height=40mm]{Legenda}
\par\end{centering}
}
\par\end{centering}
\color{black}
\centering{}\caption{{\small{}\label{fig:13}Microscale behavior, countour levels of stress
distribution in the fibers. Minimum and maximum corresponding to the
vertical color scales are: a) -2.39/+0.6 {[}}$N/mm^2${\small{}{]};
b) -2.34/+0.48 {[}}$N/mm^2${\small{}{]}; c) -6.43/-4.97 {[}}$N/mm^2${\small{}{]}.}}
\end{figure}

\par\end{center}

\begin{center}
\begin{figure}[!h]
\noindent \centering{}\includegraphics[height=75mm]{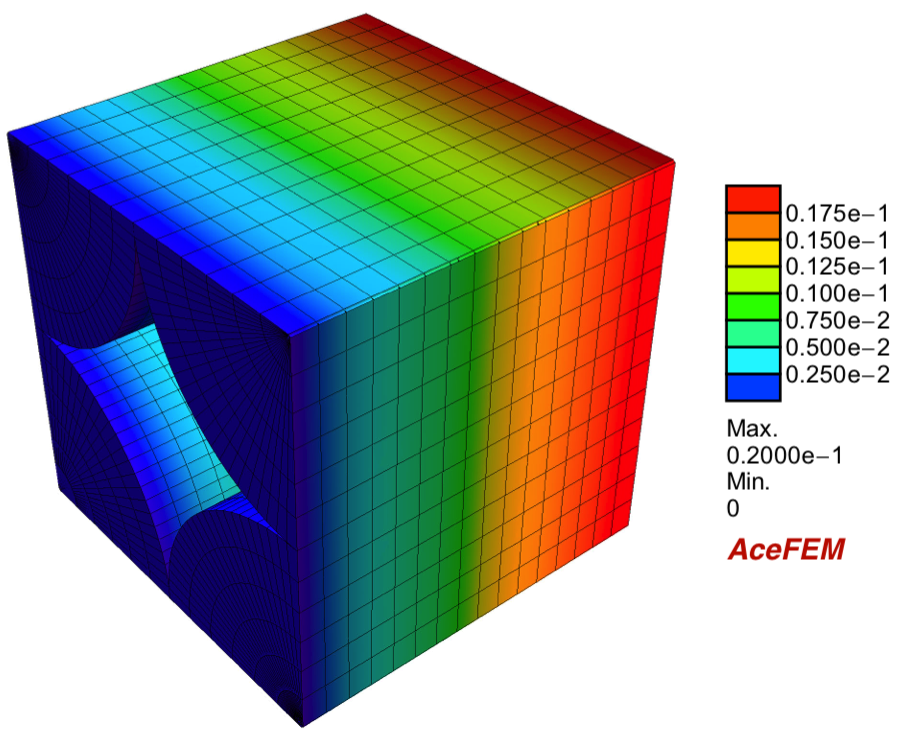}
\caption{{\small{}\label{fig:14}Electric potential distribution in the RVE.}}
\end{figure}

\par\end{center}

\begin{center}
\begin{figure}[!h]
\noindent \begin{centering}
\subfloat[i=j=11 ]{\begin{centering}
\includegraphics[height=65mm]{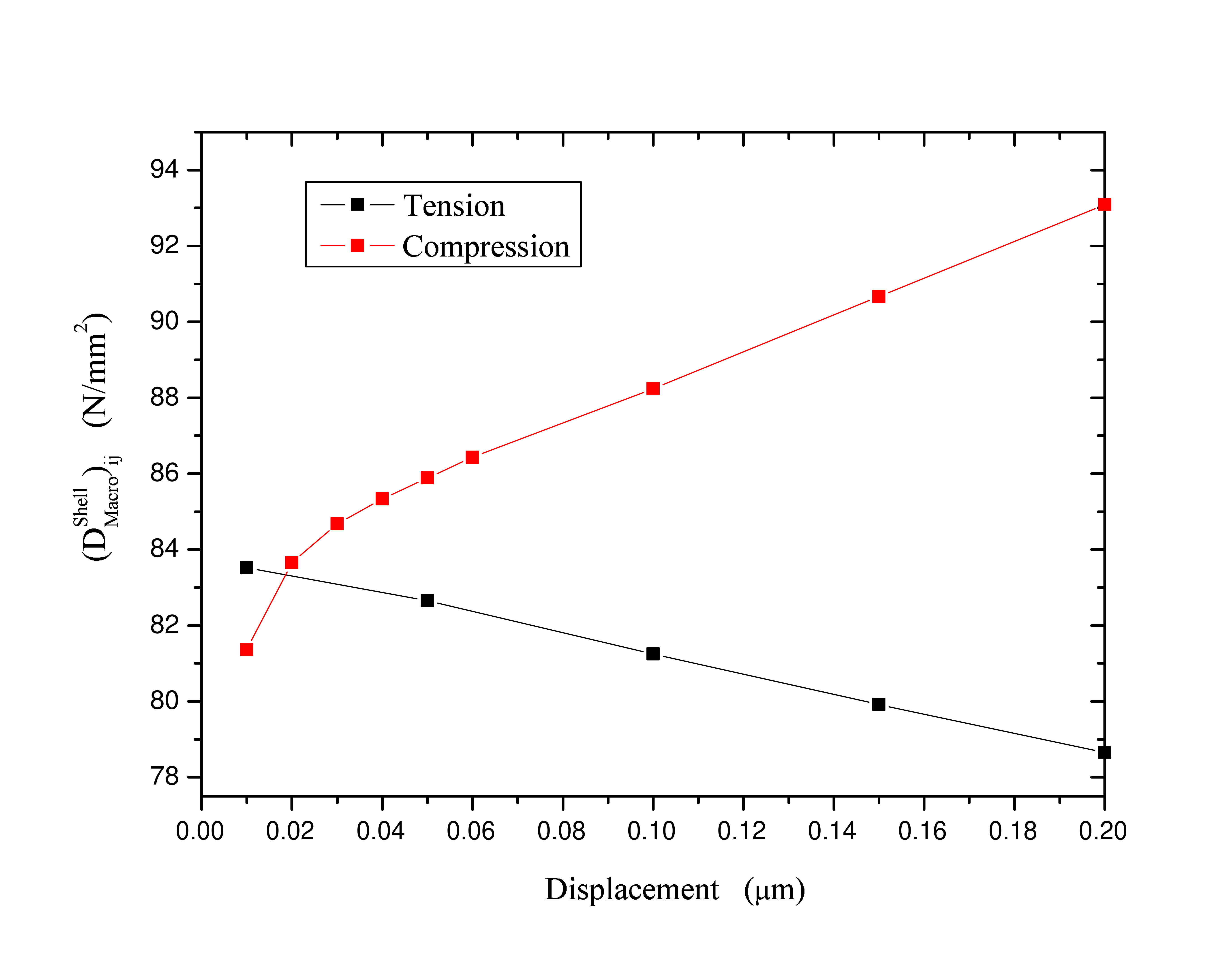}
\par\end{centering}

}\subfloat[i=1 and j=2 or i=2 and j=1 ]{\begin{centering}
\includegraphics[height=65mm]{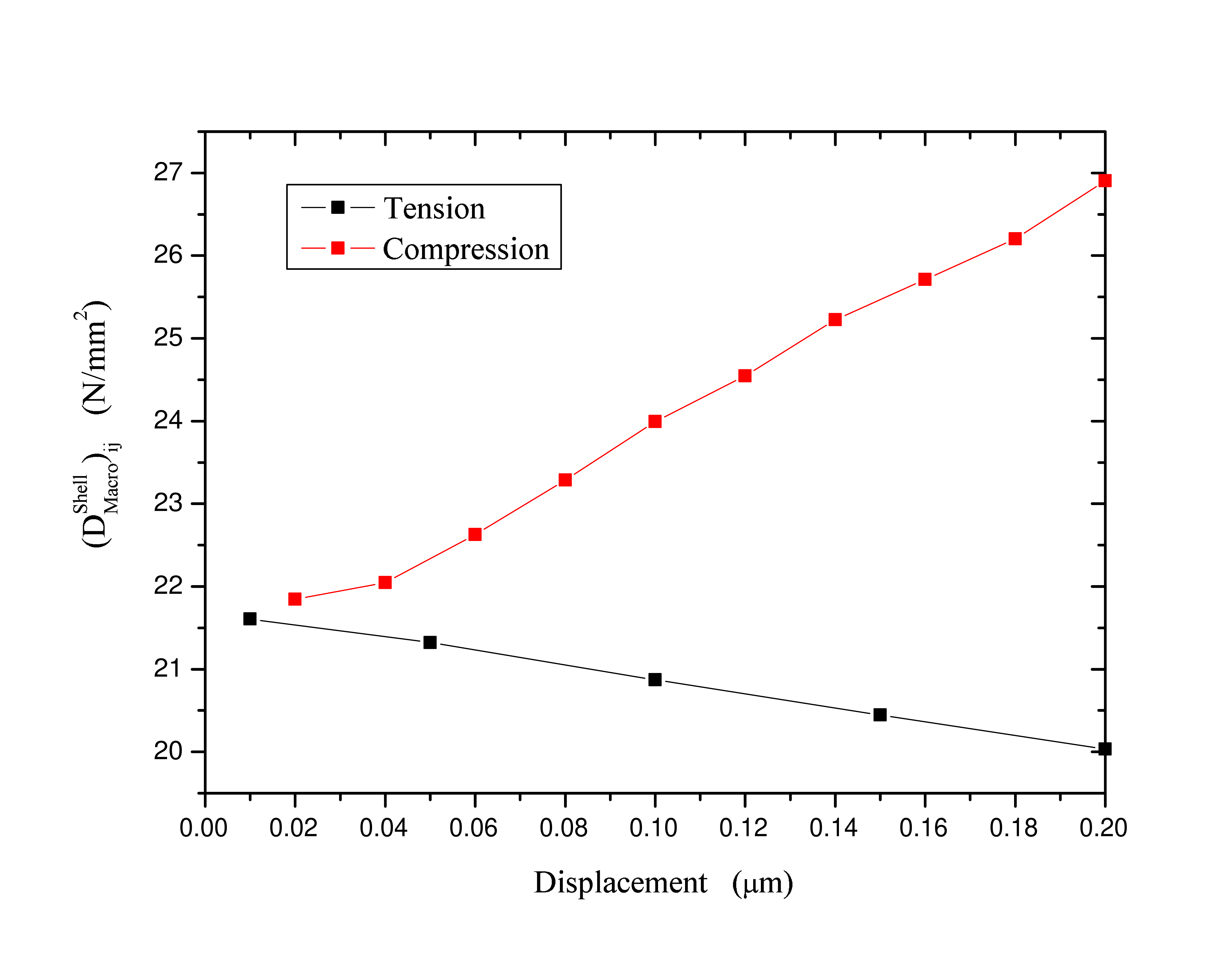}
\par\end{centering}

}
\par\end{centering}

\noindent \begin{centering}
\subfloat[i=1 and j=11]{\begin{centering}
\includegraphics[height=65mm]{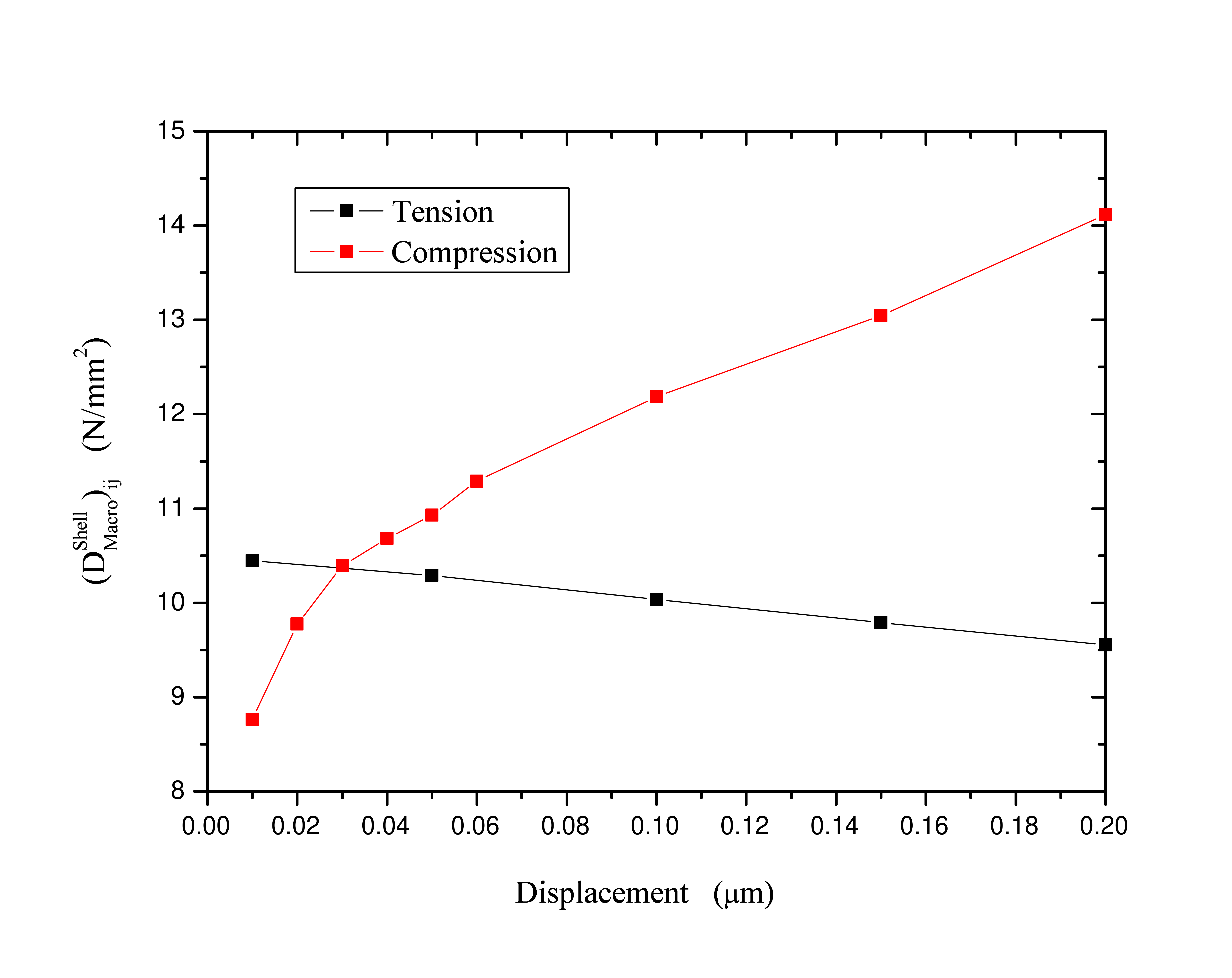}
\par\end{centering}

}\subfloat[i=1 and j=1 or i=2 and j=2]{\begin{centering}
\includegraphics[height=65mm]{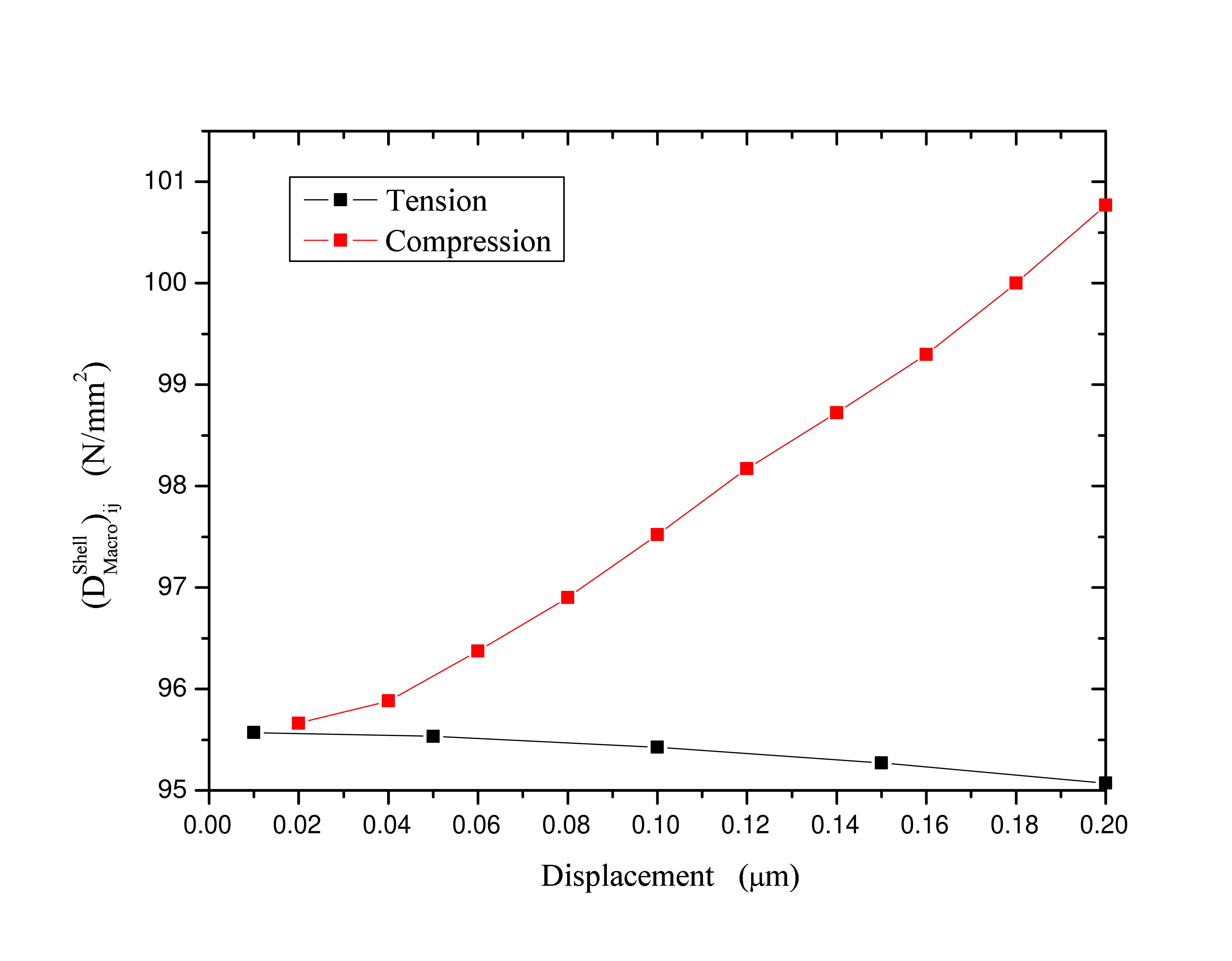}
\par\end{centering}

}
\par\end{centering}

\centering{}\caption{{\small{}\label{fig:15}Effective coefficients for the homogenized
shell.
}}
\end{figure}

\par\end{center}

\begin{thebibliography}{10}
\bibitem[1]{key-12}H. Berger, U. Gabbert, H Koeppe, R. Rodriguez-Ramos,
J. Bravo-Castillero, G. R. Diaz, J. A. Otero, G. A. Maugin. Finite
element and asymptotic homogenization methods applied to smart composite
materials. Computational Mechanics, 33, 61-67, 2003.

\bibitem[2]{key-13}H. Berger, S. Kari, U. Gabbert, R. Rodriguez-Ramos,
R. Guinovart-Diaz, J. A. Otero, J. Bravo-Castillero. An analytical
and numerical approach for calculating effective material coefficients
of piezoelectric fiber composites. Int. J. Solids Struct., 42, 5692-5714,
2004.

\bibitem[3]{key-10}J. B. Castillero, G. R. Diaz, F. J. Sabina, R.
Rodriguez-Ramos. Closed-form expressions for the effective coefficients
of fibre-reinforced composite with transversely isotropic constituents-II.
Piezoelectric and square symmetry. Mech. Mater., 33, 237-248, 2001.

\bibitem[4]{key-100001}C. Chang, V.H. Tran, J. Wang, Y. K. Fuh, L.
Lin. Direct-Write Piezoelectric Polymeric Nanogenerator with High
Energy Conversion Efficiency. Nano Lett. 10, 726\textendash{}731,
2010.

\bibitem[5]{100002}X. Chen, S. Xu, N. Yao, Y. Shi. 1.6 V Nanogenerator
for Mechanical Energy Harvesting Using PZT Nanofibers. Nano Lett.
10, 2133\textendash{}2137, 2010.

\bibitem[6]{key-17}E. W. C. Coenen, V. G. Kouznetsova, M. G. D. Geers.
Computational homogenization for heterogeneous thin sheets. Int. J.
Numer. Meth. Engng, 83, 1180-1205, 2010.

\bibitem[7]{key-23}G. Zavarise, L. De Lorenzis. The node-to-segment algorithm
for 2D frictionless contact: classical formulation and special cases. Computer Methods in Applied
Mechanics and Engineering, 198, 3428-3451, 2009.
\color{black}
\bibitem[8]{key-24}L. De Lorenzis, P. Wriggers, T. J. R. Hughes.
Isogeometric contact: a review. GAMM Mitteilungen, 37(1), 85-123, 2014.
\color{black}
\bibitem[9]{key-18}S. Fillep, J. Mergheim, P. Steinmann. Computational
modelling and homogenization of technical textiles. Eng. Structures,
50, 68-73, 2013.

\bibitem[10]{key-1}Y. Gao, Z.L. Wang. Electrostatic potential in
a bent piezoelectric nanowire. The fundamental theory of nanogenerator
and nanopiezotronics. Nano Lett. 7, 2499-2505, 2007.

\bibitem[11]{key-6}M. G. D. Geers, V. G. Kouznetsova, W. A. M. Brekelmans.
Multi-scale first-order and second-order computational homogenisation
of microstructures towards continua. Int. J. Multiscale Comput. Eng.
1, 371-386, 2003.

\bibitem[12]{key-15}S. Klinkel, W. Wagner. A piezoelectric solid
shell element based on a mixed variational formulation for geometrically
linear and nonlinear applications. Computers and Structures, 86, 38-"46,
2008.

\bibitem[13]{key-16}S. Klinkel, F. Gruttmann, W. Wagner. A mixed
shell formulation accounting for thickness strains and finite strain
3d material models. Int. J. Numerical Methods in Engineering, 75,
945-970, 2008.

\bibitem[14]{key-20}J. Korelc. Multi-language and multi-environment
generation of nonlinear finite element codes. Engineering with Computers,
18(4), 312-327, 2002.

\bibitem[15]{key-19}J. Korelc. Automation of primal and sensitivity
analysis of transient coupled problems. Computational Mechanics, 44(5),
631-649, 2009.

\bibitem[16]{key-5}V. Kouznetsova, W. A. M. Brekelmans, F. P. T.
Baaijens. An approach to micro-macro modeling of heterogeneous materials.
Comput. Mech. 27, 37-48, 2001.

\bibitem[17]{key-21}J. Lengiewicz, J. Korelc, S. Stupkiewicz. Automation
of finite element formulations for large deformation contact problems.
Int. J. Numer. Meth. Eng., 85, 1252-"1279, 2011.

\bibitem[18]{key-2002}D. Li, Y. Xia. Electrospinning of Nanofibers:
Reinventing the Wheel?. Adv. Mater., 16, 1151-1170, 2004.

\bibitem[19]{key-100003}Z. H. Liu, C. T. Pan, L. W. Lin, H. W. Lai.
Piezoelectric properties of PVDF/MWCNT nanofiber using near-field
electrospinning. Sensors and Actuators A, 193, 13\textendash{}24,
2013.

\bibitem[20]{key-100004}Z. H. Liu, C. T. Pan, L. W. Lin, J. C. Huang,
Z. Y. Ou. Direct-write PVDF nonwoven fiber fabric energy harvesters
via the hollow cylindrical near-field electrospinning process. Smart
Materials and Structures, 23, 2014.

\bibitem[21]{key-7}Miehe C. Computational micro-to-macro transitions
for discretized microstructures of heterogeneous materials at finite
strains based on the minimization of averaged incremental energy.
Comput. Methods Appl. Mech. Eng., 192, 559-591, 2003.

\bibitem[22]{key-2}L. Persano, C. Dagdeviren, Y. Su, Y. Zhang, S.
Girardo, D. Pisignano, Y. Huang, J. A. Rogers A. High performance
piezoelectric devices based on aligned arrays of nanofibers of PVDF.
Nature Communications, 1633, 4, 2013.

\bibitem[23]{key-2004}D. Pisignano. Polymer nanofibers. Cambridge:
Royal Society of Chemistry, 2013.

\bibitem[24]{key-2001}D. H. Reneker, I. Chun. Nanometre diameter
fibres of polymer, produced by electrospinning. Nanotechnology, 7,
216-223, 1996.

\bibitem[25]{key-11}F. J. Sabina, R. Rodriguez-Ramos, J. B. Castillero,
G. R. Diaz. Closed-form expressions for the effective coefficients
of fibre-reinforced composite with transversely isotropic constituents-II.
Piezoelectric and hexagonal symmetry. J. Mech. Phys. Solids, 49, 1463-79,
2001.

\bibitem[26]{key-27}J. Schroeder, M. Keip. Two-scale homogenization
of electromechanically coupled boundary value problems - Consistent
linearization and applications. Computational Mechanics, 50(2), 229-244,
2012.

\bibitem[27]{key-14}K. Schulz, S. Klinkel, W. Wagner. A finite element
formulation for piezoelectric shell structures considering geometrical
and material non-linearities. Int. J. Numer. Meth. Engng, 87, 491-520,
2011.

\bibitem[28]{key-8}P. M. Suquet. Local and global aspects in the
mathematical theory of plasticity. In Plasticity Today: Modelling,
Methods and Applications. Eds. A. Sawczuk and G. Bianchi, Elsevier
Applied Science Publishers, London, 279-310, 1985.

\bibitem[29]{key-2003}J. H. Wendorff, S. Agarwal, A. Greiner. Electrospinning:
Materials, Processing and Applications. Wiley-VCH, Weihneim, 2012.

\bibitem[30]{key-100}P. Wriggers. Computational contact mechanics.
Springer, Berlin, 2006.

\bibitem[31]{key-102}J. Yang. An introduction to the theory of piezoelectricity.
Springer Science + Business Media Inc, 2005.

\bibitem[32]{key-101}G. Zavarise, L. De Lorenzis. A modified node-to-segment
algorithm passing the contact patch test. International Journal for
Numerical Methods in Engineering, 79(4), 379-416, 2009.

\bibitem[33]{key-901}M. Bischoff, W. A. Wall, K.-U. Bletzinger and E. Ramm. Models and Finite Elements for Thin-walled Structures. Encyclopedia of Computational Mechanics. Chapter 3, 59-137, 2004.

\bibitem[34]{key-902}B. Brank, J. Korelc, A. Ibrahimbegovic. Nonlinear shell problem formulation accounting for through-the-thickness stretching and its finite element implementation. Computers and Structures, 80, 699-717, 2002.

\bibitem[35]{key-903}L. Persano, C. Dagdeviren, C. Maruccio, L. De Lorenzis, D. Pisignano. Cooperativity in the enhanced piezoelectric response of polymer nanowires. Advanced Materials, 26, 7574-7580, 2014.

\bibitem[36]{key-904}A. E. Green, P. M. Naghdi.  On the derivation of shell theories by direct approach. J. Appl. Mech., 41, 173, 1974.

\bibitem[37]{key-905}G.P. Tandon, G.J. Weng. A theory of particle-reinforced plasticity. J. Appl. Mech., 55, 126–135, 1988.

\bibitem[38]{key-906}M. Berveiller, A. Zaoui. An extension of the self-consistent scheme to plastically-flowing polycrystals. J. Mech. Phys. Solids, 26, 325–344, 1979.

\bibitem[39]{key-907}C. Gonzalez, J. Segurado, J. LLorca. Numerical simulation of elasto-plastic deformation of composites: evolution of stress microfields and implications for homogenization models. J. Mech. Phys. Solids, 52, 1573–1593, 2004.

\bibitem[40]{key-1177}P. Steinmann. Computational Nonlinear Electro-Elasticity - Getting Started. Mechanics and Electrodynamics of Magneto- and Electro-elastic Materials, CISM International Centre for Mechanical Sciences, 527, 181-230, 2011.
\color{black}

\end{thebibliography}
\end{document}